\documentclass[acmtog]{acmart}
\acmSubmissionID{1234}
\usepackage{booktabs} 
\usepackage{multirow}
\usepackage{array}
\usepackage{color}
\usepackage{pifont}
\usepackage[ruled]{algorithm2e} 

\SetAlFnt{\small}
\SetAlCapFnt{\small}
\SetAlCapNameFnt{\small}
\SetAlCapHSkip{0pt}

\acmJournal{TOG}

\newcommand{\img}{{\mathbf{I}}}

\begin{document}
\title{Invertible Tone Mapping with Selectable Styles}

\author{Zhuming Zhang*}
\affiliation{%
	\institution{The Chinese University of Hong Kong}
	\city{Hong Kong}
	\country{}
}
\author{Menghan Xia}\authornote{Equal contribution}
\affiliation{%
	\institution{The Chinese University of Hong Kong}
	\city{Hong Kong}
	\country{}
}
\author{Xueiting Liu}
\affiliation{%
\institution{Caritas Institute of Higher Education}
\city{Hong Kong}
\country{}
}
\author{Chengze Li}
\affiliation{%
\institution{The Chinese University of Hong Kong}
\city{Hong Kong}
\country{}
}
\author{Tien-Tsin Wong}\authornote{Corresponding author}
\affiliation{%
\institution{The Chinese University of Hong Kong}
\city{Hong Kong}
\country{}
}
\email{ttwong@cse.cuhk.edu.hk}

\begin{teaserfigure}
	\centering
	\includegraphics[width=0.9\linewidth]{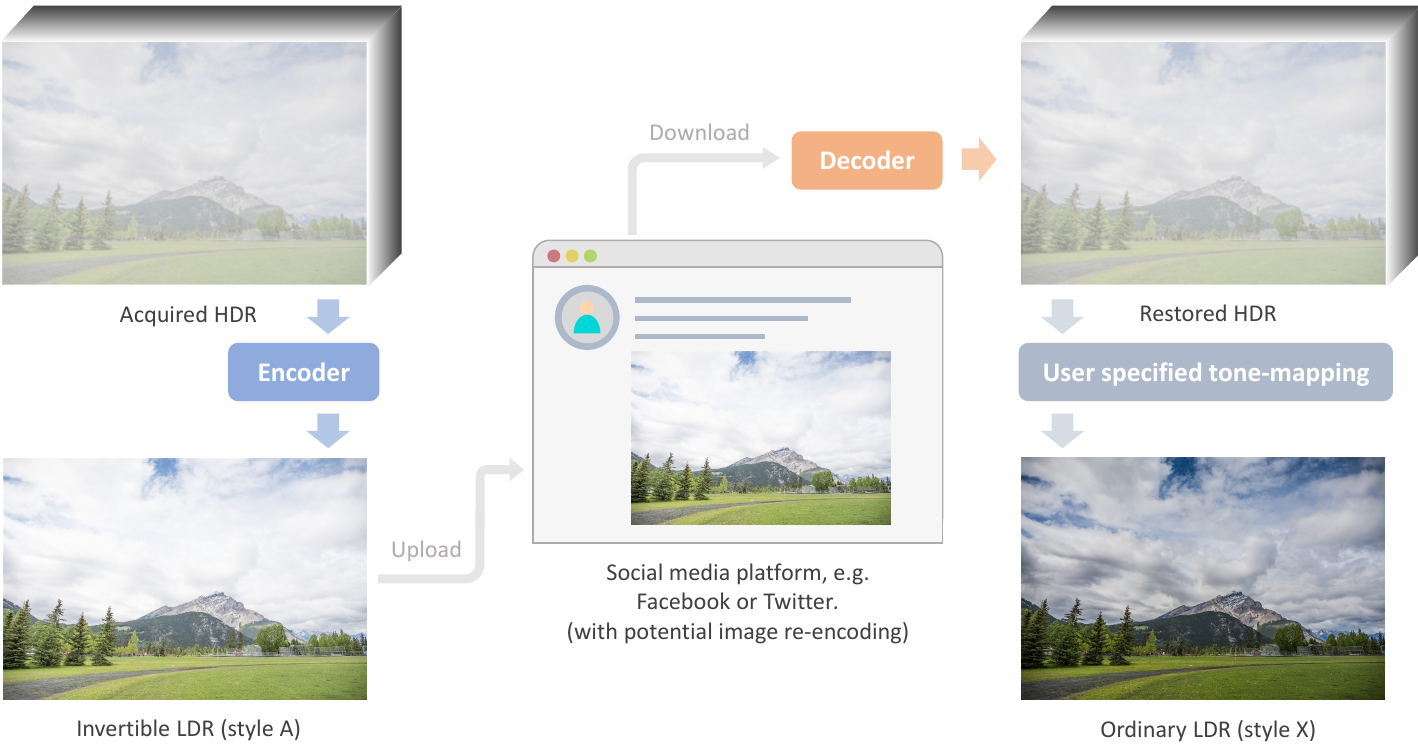}	
	\caption{Once the HDR image is acquired by the camera CMOS/CCD, our encoder converts it into an LDR image that is with 8-bit color depth and can be shared on social media platform, e.g. \textit{Facebook}, \textit{Twitter}, etc. Whenever the original HDR is needed for other usages, e.g., tone mapping with user preferred style, or visualization on HDR display, it can be restored from the invertible LDR image through our decoder.}
	\label{fig:teaser}
\end{teaserfigure}

\begin{abstract}
Although digital cameras can acquire high-dynamic range (HDR) images, the captured HDR information are mostly quantized to low-dynamic range (LDR) images for display compatibility and compact storage. In this paper, we propose an invertible tone mapping method that converts the multi-exposure HDR to a true LDR (8-bit per color channel) and reserves the capability to accurately restore the original HDR from this {\em invertible LDR}. Our invertible LDR can mimic the appearance of a user-selected tone mapping style. It can be shared over any existing social network platforms that may re-encode or format-convert the uploaded images, without much hurting the accuracy of the restored HDR counterpart. 
To achieve this, we regard the tone mapping and the restoration as coupled processes, and formulate them as an encoding-and-decoding problem through convolutional neural networks. Particularly, our model supports pluggable style modulators, each of which bakes a specific tone mapping style, and thus favors the application flexibility.
Our method is evaluated with a rich variety of HDR images and multiple tone mapping operators, which shows the superiority over relevant state-of-the-art methods.
Also, we conduct ablation study to justify our design and discuss the robustness and generality toward real applications.
\end{abstract}

%
%



\maketitle

\section{Introduction}
\label{sec:introduction}

\begin{figure}[!t]
	\centering
	\includegraphics[width=1.0\linewidth]{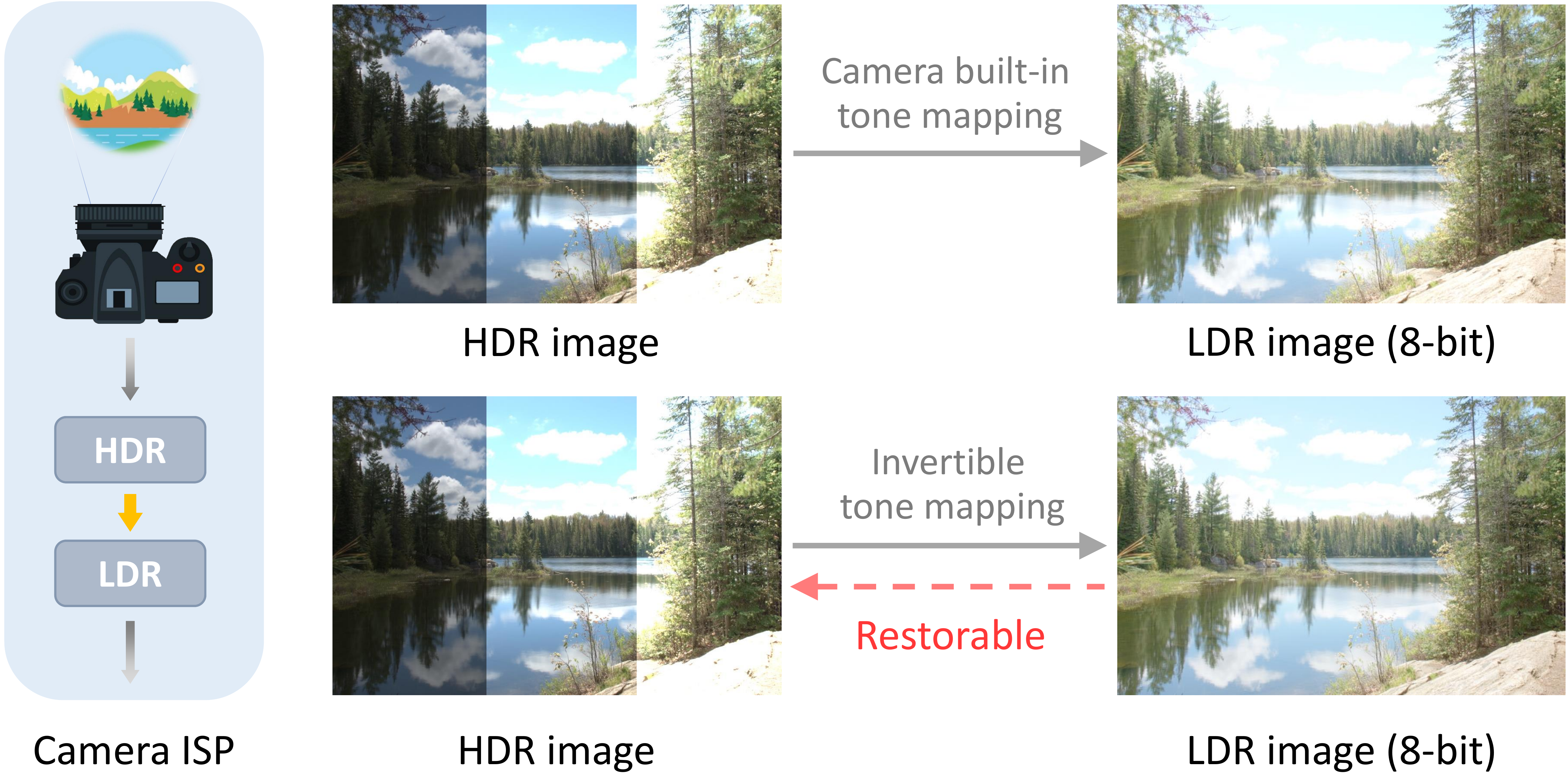}\vspace{-0.5em}	
	\caption{Tone mapping module of digital imaging pipeline: existing solution (upper row) vs. our proposed one (bottom row). In this figure, the HDR image is represented as three concatenated strips in three different exposures, as we cannot visualize the full dynamic range on paper. }\vspace{-1.0em}
	\label{fig:intro_camera_flow}
\end{figure}

Modern digital cameras can acquire high-dynamic range (HDR) images with a precision of 14 bits or 16 bits per color channel at single exposure. With multiple images acquired at multiple exposures, HDR with even higher precision can be obtained by merging them~\cite{debevec1997recovering, reinhard2010high, banterle2017advanced}.
However, these HDR images are mostly immediately quantized (or tone-mapped) to low-dynamic range (LDR) images, not solely for compact storage, but more importantly for compatible display and distribution (e.g. sharing online), as LDR displays are still dominant today. Such quantization permanently removes visual details from the acquired HDR, e.g. the over-exposed sky in Fig.~\ref{fig:intro_camera_flow}(upper right). 

It is desirable to recover the HDR information from a tone-mapped LDR images, since HDR significantly favors further editing, tone-mapping to another LDR style, or other analysis purpose.
Several inverse tone mapping techniques~\cite{endo2017deep,eilertsen2017hdr,marnerides2018expandnet} are hence proposed to infer the HDR from a normal LDR image. Nevertheless, such inverse problem is ill-posed by nature, which makes it intractable to accurately recover the original HDR. 
Another approach is to split the HDR information into an LDR compatible part and an incompatible auxiliary part that stores the residual information, e.g. double-layer HDR compression (such as the latest JPEG-XT)~\cite{ward2005jpeg,Alessandro2016jpegxt}. However, these approaches produce no independent LDR bitmap but meta data (bitmap $+$ auxiliary) that is not truly LDR compatible. It requires a tailor-made image format (e.g. JPEG-XT), and the auxiliary data may get lost once the image is re-encoded or format-converted, e.g. the auxiliary data of image files gets trimmed away when uploaded to social media platform \textit{Facebook}. HDR companding~\cite{li2005compressing} determines a pair of carefully designed symmetric operations, i.e. tone mapping and inverse tone mapping, so as to facilitate the HDR recovery from the tone-mapped LDR. However, the generated LDR style is fixed and it can hardly be extended to other LDR styles. Furthermore, the hand-crafted data representation makes the restoration accuracy limited. Table.~\ref{table:feature_comp} compares the features of these techniques.

In this paper, we propose a novel solution, named as {\em invertible tone mapping}, to produce the true LDRs that mimic the visual appearance of the user-desired tone mapping styles and reserves the restorability to the original HDR. 
Fig.~\ref{fig:teaser} demonstrates an example that an HDR captured by SLR camera is converted to {\em invertible LDR} with selected style for distribution over social media platforms. Even with the potential re-encoding and reformat conversion, this shared copy of LDR can still be restored to the original HDR variant for further processing, e.g. tone-mapping to another LDR style. Note that, the hardly observable cloud textures in our invertible LDR (left bottom) becomes apparent in the re-tonemapped LDR (right bottom).
To achieve this goal, we consider the tone mapping and restoration as a whole within an encoding and decoding framework, as inspired by invertible grayscale~\cite{xia-2018-invertible}. Unfortunately, directly applying the invertible network~\cite{xia-2018-invertible} does not work, because of two facts: (i) Tone mapping is more tricky than converting color to grayscale, which requires to consider both global and local informations; (ii) Controllable tone mapping style should be enabled by the tone mapping network. Hence, we propose a dual-branch network architecture that utilize global and local receptive fields comprehensively. To allow selectable tone mapping styles, our model support pluggable style modulators that bake various styles respectively, and they can further be extended to perform parametric control for each style with additional parameter input. 
Our model is trained with the loss function consisting of two loss terms, the invertibility loss and the style loss. The invertibility loss penalizes the difference between the restored and the original HDRs, and the style loss encourages the similarity between the generated LDR and the target tone mapping appearance.  To make the encoding scheme tolerant to lossy compression, e.g. JPEG, we introduce a JPEG layer before the decoding network during training and consequently our {\em invertible LDR} is highly resistant to the common JPEG noise. 

\begin{table}[!t]
	\vspace{0.1in}
	\centering
	\renewcommand{\tabcolsep}{2pt}
	\caption{Feature comparison among LDR-to-HDR recovery techniques.}
	\vspace{-0.1in}
	\begin{tabular}{ccccc}
		\hline
		\multirow{2}{*}{Method}  & \multicolumn{3}{c}{Input LDR}            & \multicolumn{1}{c}{HDR}\\
		 \cline{2-4}                          & Bitmap       & Re-encodable              &  Multi-style           & Accuracy   \\ \hline
		HDR Compression              & {\color{red}\ding{55}}        & {\color{red}\ding{55}}         & \textcolor[rgb]{0,0.75,0}{\ding{51}}      & \textcolor[rgb]{0,0.75,0}{\textit{high}}          \\
		Inverse Tone-map		      & \textcolor[rgb]{0,0.75,0}{\ding{51}}    & \textcolor[rgb]{0,0.75,0}{\ding{51}}     & \textcolor[rgb]{0,0.75,0}{\ding{51}}      &  {\color{red}\textit{low}}          \\
		HDR Companding		         & \textcolor[rgb]{0,0.75,0}{\ding{51}}    & \textcolor[rgb]{0,0.75,0}{\ding{51}}     & {\color{red}\ding{55}}          &  {\color{red}\textit{limited}}    \\
		Ours		                              & \textcolor[rgb]{0,0.75,0}{\ding{51}}    & \textcolor[rgb]{0,0.75,0}{\ding{51}}     & \textcolor[rgb]{0,0.75,0}{\ding{51}}      & \textcolor[rgb]{0,0.75,0}{\textit{high}}           \\
		\hline
	\end{tabular}\vspace{-1.0em}
	\label{table:feature_comp}
\end{table}

To validate our model, we tested it over a large set of HDR images of different categories and dynamic ranges. Various existing tone mapping operators are adopted to demonstrate its generality in mimicking various appearance styles. 
We also quantitatively compared our method to existing methods such as HDR companding, inverse tone mapping, and JPEG-XT standard. Experiments show that our method significantly outperforms all these competitors in terms of HDR restoration accuracy and functional flexibility.
Our contributions can be summarized as:

\begin{itemize}
	\item We propose an invertible tone mapping method that enables an HDR image to be accurately represented as an ordinary LDR bitmap with controllable tone-mapping style. To our best knowledge, it is the first technique to achieve this non-trivial feature.
	\item We present a parametric invertible generative model that allows multiple style modulators as plug-in to be trained through a single model. The incremental training scheme eases the memory or computation burden of adding new styles.
	\item Our proposed technique possesses powerful application potentials and may trigger new market demands. For example, it can be embedded into the camera pipeline, or serve toward platform-compatible HDR representation.
\end{itemize}
\section{Related works}
\label{sec:related_works}

\subsection{Tone Mapping}
\label{subsec:tone_mapping}

\paragraph{Global Operators}Tone mapping operators have been proposed to quantize HDR to LDR for better compatibility and visualization on LDR displays. Comprehensive surveys have been made by Reinhard et al.~\shortcite{reinhard2010high} and Banterle et al.~\shortcite{banterle2017advanced}.
Global operators compress the dynamic range of an HDR image in a spatially invariant manner~\cite{reinhard2002photographic, tumblin1993tone, ward1994contrast, drago2003adaptive, reinhard2005dynamic}.
That is, the same compression curve is applied over the whole image.
In particular, Tumblin et al.~\shortcite{tumblin1993tone} proposed to produce the brightness of the generated LDR image as close to the real-world sensation as possible. Ward~\shortcite{ward1994contrast} proposed to determine the brightness and contrast in the tone-mapped LDR image based on psychophysics studies.
Inspired by photographic practices, Reinhard et al.~\shortcite{reinhard2002photographic} proposed a photographic tone reproduction technique to solve the HDR-to-LDR problem.
Drago et al.~\shortcite{drago2003adaptive} proposed to logarithmically compress the luminance values, imitating the human response to light.

\paragraph{Local Operators}
Comparing to global operators, local operators compress HDR images in a spatially variant manner~\cite{durand2002fast,farbman2008edge,paris2011local,aubry2014fast,fattal2002gradient}.
They usually decompose the HDR into layers, adjust each layer independently, and recombine them into the final LDR.
In particular, Durand and Dorsey~\shortcite{durand2002fast} proposed to decompose the HDR into a base layer and a detail layer, and then produce the LDR by compressing the base layer. Fattal et al.~\shortcite{fattal2002gradient} proposed to compress the dynamic range by attenuating the magnitudes of large gradients at different scales.
Different from the methods above, Li et al.~\shortcite{li2005compressing} aimed at the invertibility of tone mapping, which is similar to ours. They proposed to use a symmetric analysis-synthesis filter bank to decompose the image into multiple subbands. Both forward tone mapping (dynamic range compression) and inverse tone mapping (dynamic range expansion) can be achieved by applying the local gain control to the subbands. However, their method suffers from insufficient HDR accuracy. Besides, it can only produce a fixed tone mapping style.  

While all existing tone mapping operators aim at generating LDR images with enhanced visual richness, visual information may still be lost because their primary concern is the visual appearance, instead of information preservation. 
In sharp contrast, our invertible tone mapping can generate LDR such that its HDR counterpart can be accurately restored.

\subsection{Inverse Tone Mapping}
\label{subsec:inverse_tone_mapping}

\paragraph{Procedural Methods.}
An in-depth survey of non-data-driven inverse tone mapping methods can be found in~\cite{banterle2009high}.
To infer HDR from LDR, researchers~\cite{banterle2006inverse,rempel2007ldr2hdr,kovaleski2014high} expanded the LDR by inverting the global tone mapping operator using various expand maps~\cite{reinhard2002photographic}.
Huo et al.~\shortcite{huo2014physiological} proposed to recover the luminance channel of the HDR based on physiological studies.
Wang et al.~\shortcite{wang2015pseudo} proposed to segment the LDR images into regions where each region is enhanced independently.
While the above methods can produce reasonable global contrast in the estimated HDR, they have very strong assumptions on the tone mapping operator and generally can only be applied on global tone mapping operators. What's more, the missing details cannot be recovered.

\paragraph{Data-Driven Methods.}
Recently, with the development of convolutional neural networks (CNNs), a couple of data-driven inverse tone mapping methods have also been proposed.
Zhang et al.~\shortcite{zhang2017learning} proposed to infer the HDR from a daytime outdoor panorama LDR image.
Eilertsen et al.~\shortcite{eilertsen2017hdr} proposed to separate the input LDR image into saturated and non-saturated regions. The content in the saturated regions is inferred by the network, while that in the non-saturated regions is processed with a fixed inverse camera response curve.
Endo et al.~\shortcite{endo2017deep} proposed to first synthesize multiple LDR images of different exposures and then merge them into a single HDR image using standard merging algorithms.
Most recently, Marnerides et al.~\shortcite{marnerides2018expandnet} proposed to predict HDR from LDR with a CNN trained on paired HDR and LDR images generated with different tone mapping operators. 

However, inverse tone mapping by itself is an ill-posed problem because the information is permanently lost during the forward tone mapping. Therefore, even though the above methods can infer visually reasonable HDR, the inferred content cannot be guaranteed to conform to the original HDR. Instead of solving this ill-posed inverse tone mapping as an isolated problem, we regard the forward tone mapping and inverse tone mapping as a whole, and design an invertible tone mapping that allows {\em HDR restoration}, instead of estimation or inference.

\begin{figure*}[!t]
	\centering
	\includegraphics[width=1.0\linewidth]{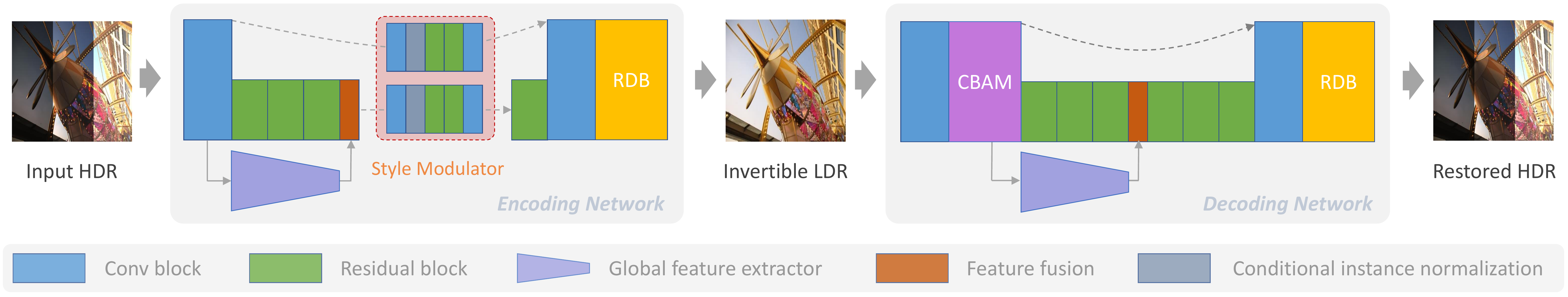}
	\caption{System Overview. The encoding network converts an HDR input into an LDR, which can control the generative styles by adopting different style modulators. Oppositely, the decoding network reconstructs the original HDR variant from the generated LDR. CBAM is a convolutional attention module. RDB is a residual dense block.}\vspace{-0.5em}
	\label{fig:overview}
\end{figure*}

\subsection{HDR Compression}
\label{subsec:HDR_compression}

Even though compression is not our primary goal, our invertible LDR offers a compact storage as a side-product. 
Existing HDR compression methods can be categorized into single-layer and double-layer methods. Single-layer methods solely aim at the compression ratio and ignore the backward compatibility to LDR display or coding standard~\cite{larson1998logluv, mantiuk2004perception, miller2013perceptual}. Their compressed data is not visualizable. In sharp contrast, our invertible LDR is a true LDR. It is visualizable on any existing LDR display, and fully compatible with any existing image coding standard.

The double-layer HDR compression methods~\cite{spaulding2003using, ward2005jpeg, mantiuk2006backward, artusi2016jpeg} decompose the HDR into a base LDR layer and an auxiliary layer, in which the LDR is stored in normal JPEG and the residual layer is stored as the metadata of JPEG. When the compressed image is decoded/displayed by legacy codecs, only the LDR is extracted and the metadata is simply ignored. In other words, the generated data is not solely an LDR, but an (LDR + auxiliary) data. Hence, a tailor-made format (e.g. JPEG-XT) and special treatment (that avoids damaging the auxiliary data) are needed. Otherwise, converting this (LDR + auxiliary) data to other image formats for online sharing or distribution may immediately trim away the auxiliary data, and lead to permanent loss of the HDR. In sharp contrast, our invertible LDR is a true LDR, and hence is fully compatible to any existing LDR platforms. File format conversion leads to zero or insignificant damage to our HDR recovery.

\section{Overview}
\label{sec:overview}

Fig.~\ref{fig:overview} shows our framework that consists of tone mapping path via the encoding network and a restoration path via the decoding network. Particularly, the tone mapping converts an input HDR image $\img_h$ to the invertible LDR image $\img_l$ (an standard 8-bit per color channel RGB image), while the restoration takes the generated $\img_l$ to recover an HDR $\hat{\img}_h$. They are accomplished via an encoding network $E$ and a decoding network $D$:
\begin{eqnarray}
\label{Eq:fomula}
\img_l &= &E(\img_{\rm h}|\mathcal{W}_E) ,\\
\tilde{\img}_{\rm h}& = & D(\img_l|\mathcal{W}_{D}), 
\end{eqnarray}
where $\mathcal{W}_E$ and $\mathcal{W}_D$ are the weights of networks $E$ and $D$ respectively. $E$ and $D$ share the same backbone architecture that has a overall U-shaped structure and a pyramidal-downscaling global branch. In particular, the design of pluggable style modulators enables our model to introduce new tone mapping styles efficiently. The structure design is introduced in Section~\ref{subsec:network_structure}.
There are two goals for the invertible LDR generation. First, the restored HDR $\tilde{\img}_{\rm h}$ should be as close as possible to the original HDR ${\img}_{\rm h}$. Second, the LDR $\img_{\rm l}$ should have a similar appearance to the target LDR $\hat{\img}_{\rm l}$, tone-mapped from the original HDR with certain tone mapping operators. 
Accordingly, the loss function with two terms, i.e. invertibility loss and style loss, is formulated to achieve these goals (detailed in Section~\ref{subsec:loss_func}). The two networks are jointly trained to learn the optimal encoding/decoding scheme through supervised training. 
For practical use, our model can be optionally trained with an additional JPEG layer to equip the invertible LDR with tolerance to JPEG compression noise.

The HDR intensity range varies largely and it poses challenge to the generality of CNN model. Thus, we propose to map the input HDR to a normalized domain before feeding to the network. In particular, we convert all input HDR (3-channel RGB images with 14-bit or 16-bit precision) to the LUV color space and stored as 3-channel LUV images with 32-bit precision. Following the formula of~\cite{Ward1998}, we first convert the pixel values from the RGB space to the XYZ color space. Then we compute the $LUV$ values by applying the normalization as
\begin{equation}\label{Eq:weight}
\left\{
\begin{aligned}
L &= N(\log(Y)) \\
U &= \frac{2.48X}{81(X+15Y+3Z)} \\
V &= \frac{0.62Y}{9(X+15Y+3Z)}
\end{aligned}
\right.
\end{equation}
Here, $N(\cdot)$ means linearly rescaling to the range of $[0,1]$. Thanks to the normalization, the pixel values of an HDR image are generally distributed in a compact manner (in the range of $[0,1]$ for all channels), which eases the training difficulty.
Accordingly, the HDR image restored by the decoding network will be further mapped back to its original value range in RGB color space.
\section{Approach}
\label{sec:approach}

\subsection{Network Architecture}
\label{subsec:network_structure}

Our method works with an encoding/decoding scheme, so we adopt the same backbone for the encoding network and decoding network to favor symmetric learning.
Following the same spirit of~\cite{Gharbi2017,marnerides2018expandnet}, our backbone network has a global branch aside with the main branch, so as to utilize the global and local information comprehensively. The rationale is that the global contrast and local details are two balancing ingredients of tone mapping algorithms and they are also crucial clues to infer the HDR information. The conceptual diagram is illustrated in Fig.~\ref{fig:overview}.
The global branch starts with a spatial pooling layer~\cite{He2015SPP} that downsamples the input feature maps with fixed target size (i.e. $32\times 32$) and pass them to the pyramidal-downscaling layers. To preserve the key statistic data, both the maximal and mean values are sampled in each sampling unit. The global feature vector is further replicated spatially and fused with the main branch features along channels. 
Besides, the computation power is mainly enabled by the cascaded residual blocks in the middle and a residual dense block~\cite{Zhang2018} at the rear to exploit multi-level features. 

Inspired by~\cite{ChenDD2017}, our encoding network is equipped with style modulators as plug-in, which determine the styles of the invertible LDR. This design offers the flexibility to allow any number of tone mapping styles being learned through a single model, where the shared backbone takes charge of remapping the input HDR into a compact range with extra information embedded while the style modulator performs the style transfer in feature domain. 
For the sake of the application convenience, the decoding network has no explicit individual modules to handle input LDRs of different styles. Instead, we employ a single network to restore the original HDR from various LDR variants, where the CBAM~\cite{Woo2018} is utilized to generate dynamic channel and spatial attentions conditioning on the input features. This module works well for the decoding network to handle multiple LDR styles in an adaptive manner.

\paragraph{Quantization.}
To save the generated LDR as common bitmap format (i.e. 8 bits for each R, G, and B channels), we adopt a quantization module that converts floating-point values to 8-bit unsigned int. Here, we simply use rounding operation as the quantization module in the forwarding process, and the backward gradient is calculated through Straight-Through Estimator~\cite{Bengio2013}. It enables the encoding scheme to resist the quantization error and hence preserves the embedded information.

\subsection{Pluggable Style Modulator }
\label{subsec:style_modulator}

%
\begin{figure}[!t]
	\centering
	\includegraphics[width=1.0\linewidth]{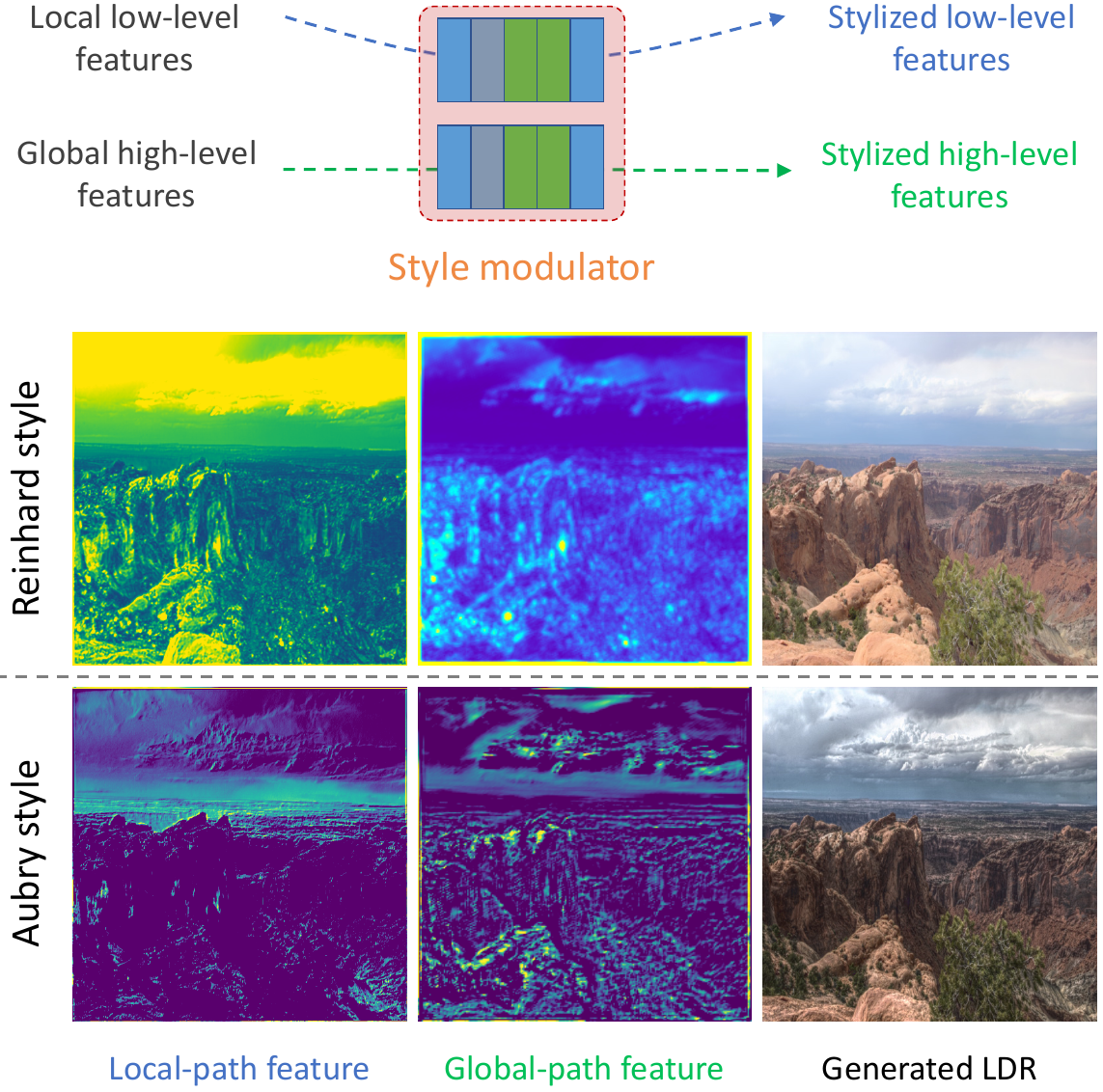}
	\caption{Feature visualization for dual-path style modulator. Two tone mapping styles, a global operator~\cite{reinhard2002photographic} and a local operator~\cite{aubry2014fast}, are taken for comparison.}
	\label{fig:style_feature}
\end{figure}

The style modulator consists of two paths, one works on low-level local information and the other works on high-level global features. This design is motivated by the fact that existing tone mapping styles mainly vary at the balance of global contrast and local details. Besides, inspired by the efficiency of conditional instance normalization~\cite{Dumoulin2017} in style transfer, we adopt this design in both paths of our style modulator, so as to facilitate the style learning.
Fig.~\ref{fig:style_feature} visualizes the feature maps generated by two style modulators that are associated with different tone-mapping styles. For the global tone-mapping style, the global path mainly controls the global illumination and contrast while the local path works on concrete details. For the local tone-mapping style, the global path not only manages the global contrast but also tunes the local contrast, while the local path focuses more on gradient variation or edge features. So, with the pluggable style modulators, multiple tone-mapping styles can be learned by our model. It equips our model with selectable tone-mapping styles while only introduces slight overhead in memory and storage.

In fact, there are two optional strategies for multi-style training: (i) Jointly training all the desired style modulators along with the backbone network; (ii) Training the backbone network with one or several style modulators at first and then incrementally training other newly introduced ones with the pretrained backbone network fixed. In theory, it is suboptimal to incrementally train the style modulators, which may has a performance inferior to the jointing training strategy. To inspect the difference, we conduct an ablation study on the alternatives in Section~\ref{subsec:ablation_study}. Anyhow, the incremental training still has its unique advantages, i.e. the flexibility to efficiently add new tone-mapping styles to a trained model, which is especially appealing to practical applications.


\subsection{Loss Function}
\label{subsec:loss_func}

The generated invertible LDR is required to have two properties: on the one hand, it should be restorable to the HDR input via the decoding network; on the other,  it should be visually similar to the target style. Accordingly, our loss function consist of two terms, the invertibility loss $\mathcal{L}_{\rm inv}$ and the style loss $\mathcal{L}_{\rm sty}$, which are respectively introduced below.

\paragraph{Invertibility Loss.}
The invertibility loss $\mathcal{L}_{\rm inv}$ measures how close the restored HDR $\hat{\img}_{\rm h}$ is to the original HDR input $\img_{\rm h}$. Specifically, $\mathcal{L}_{\rm inv}$ is defined as
\begin{equation}
\label{eq:inv_loss}
\mathcal{L}_{\rm inv} =  ||\tilde{\img}_{\rm h}-\img_{\rm h}||_2 + \sigma {\rm SSIM}(\tilde{\img}_h, \img_{\rm h}),
\end{equation}
where ${\rm SSIM}(\cdot)$ denotes the structural similarity index measure~\cite{Wang2004}, which monitors the perceived change in structural information, as complement with the pixelwise measurement of $L_2$. $\beta=5.0e-2$ are empirically used to balance magnitudes when the pixel values ranges in $[0,1]$.

\begin{figure}[!t]
	\centering
	\includegraphics[width=1.0\linewidth]{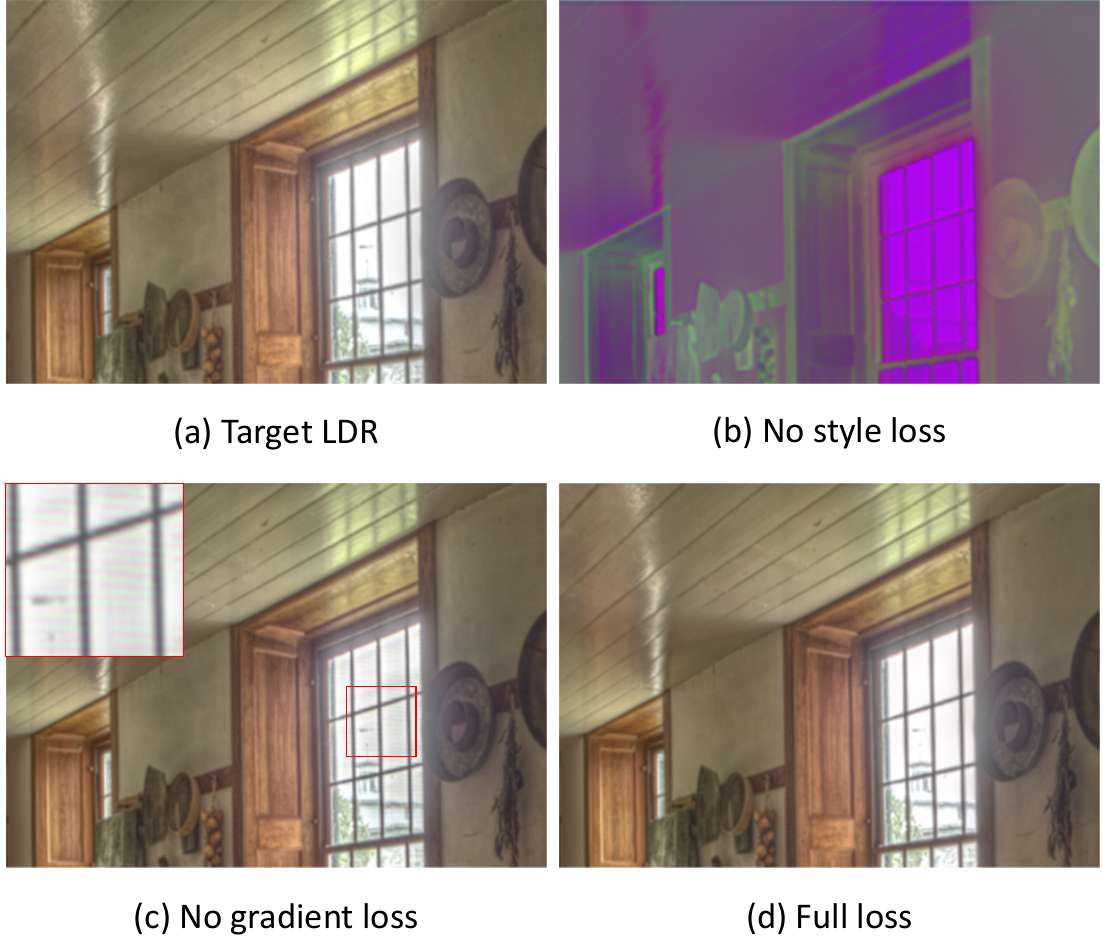}
	\caption{Ablation on the style loss ingredient. The target LDR is tone mapped by the local tone-mapping operator~\cite{aubry2014fast}.}
	\label{fig:ablation_loss}
\end{figure}

\paragraph{Style Loss.}
We first employ $L_2$-norm, $l_{\rm pix}=||\img_{\rm l}-\hat{\img}_{\rm l}||_2$, to measure the deviation between the generated LDR and the target LDR, and find that some extra structural pattern may be introduced (see Fig.\ref{fig:ablation_loss}(c)). It is an adaptive encoding phenomena driven by the invertibility loss, since the input HDR need to be represented by the LDR. To suppress such gradient artifacts, we further calculate the $L_2$-norm in the gradient domain, $l_{\rm grad}=||\nabla\img_{\rm l}-\nabla\hat{\img}_{\rm l}||_2$, which is more sensitive to gradient inconsistency, as justified by the result shown in Fig.~\ref{fig:ablation_loss}(d). However, some color deviation still exist even under the constraint of these two terms. So, we also impose the perceptual similarity in the feature space of a pretrained classification model, $l_{\rm perc}=||\Phi_i(\img_{\rm l})-\Phi_i(\hat{\img}_{\rm l})||_2$, where $\Phi(\cdot)_i$ denotes the feature maps of the i-th layer of the pretrained VGG-19~\cite{Simonyan2015}. Then, an invertible LDR that is visually similar to the target LDR can be achieved by the combined style loss:
\begin{equation}
\label{eq:style_loss}
\mathcal{L}_{\rm sty} = ||\img_{\rm l}-\hat{\img}_{\rm l}||_2 + \alpha ||\nabla\img_{\rm l}-\nabla\hat{\img}_{\rm l}||_2 + \beta ||\Phi_i(\img_{\rm l})-\Phi_i(\hat{\img}_{\rm l})||_2,
\end{equation}
where $\alpha=0.1$ and $\beta=1.0e-3$ are empirically used to balance the loss magnitudes of different terms. In particular, we use the feature maps of the 28-th layer of VGG-19, "conv4-4", to capture perceptual property. 

According to the formula in Eq.~\ref{Eq:fomula}, the invertibility loss $\mathcal{L}_{\rm inv}$ back-propagates gradient to both the encoding network and the decoding network, while the style loss $\mathcal{L}_{\rm sty}$ only back-propagates gradient to the encoding network. Hence, the overall loss function can be depicted as the linear combination of the two terms:
\begin{equation}
\label{eqn:total_loss}
\mathcal{L} = \lambda \mathcal{L}_{\rm inv}(\mathcal{W}_{E},\mathcal{W}_{D}))+(1-\lambda)\mathcal{L}_{\rm sty}(\mathcal{W}_{E}),
\end{equation}
where the hyper-parameter $\lambda$ balances between the HDR restoration accuracy and the LDR visual similarity to the target style. To be specific, raising $\lambda$ relaxes the regularization on LDR style and enhances HDR encoding accuracy. As default, we use $\lambda=0.5$ unless specially mentioned.

\subsection{Dataset Preparation and Training Details}
\label{subsec:training}

\paragraph{Dataset Preparation}
Since the multi-exposure HDR images (HDR by merging images at different exposure levels)~\cite{debevec1997recovering, reinhard2010high, banterle2017advanced} have the widest dynamic range (comparing to single-exposure ones) and hence are more challenging, we focus on multi-exposure HDR images in our experiments. We collected 677 HDR images from public available dataset, including Stanford dataset~\cite{xiao2002high}, the dataset attached to the book authored by Reinhard et al.~\shortcite{reinhard2010high}, the Funt dataset~\cite{funt2010rehabilitation}, HDR-Eye dataset~\cite{korshunov2014crowdsourcing} and the Fairchild dataset~\cite{fairchild2007hdr}. To evaluate the generality of our model, we intentionally use the challenging HDR-Eye and Fiarchild dataset as out evaluation dataset while use the rest for training, instead of randomly splitting the mixed dataset. So, there are 530 HDR image for training and 147 HDR image for testing.
As tone mapping depends on global information of the image, we can not augment the training dataset by cropping local patches that has a different data distribution from full-range images. Instead, we first resize each HDR image to $640 \times 640$ and randomly crop 10 patches of $512 \times 512$ with tone randomly tuned. Each patch is further augmented by horizontal flipping. Here no vertical flipping is performed due to the sky location prior. In total, our training set contains 10,600 HDR image samples.

To test our method in mimicking different tone mapping styles, four well-known tone mapping operators are selected, including one global operator \cite{reinhard2002photographic} and three local operators \cite{durand2002fast, li2005compressing, aubry2014fast}. The Reinhard operator~\shortcite{reinhard2002photographic} simulates the photographic tone reproduction technique. The Durand operator~\shortcite{durand2002fast} optimizes the global contrast of the whole image while preserving the local details. The Li operator~\shortcite{li2005compressing} compresses the dynamic range in a multi-scale manner with invertibility consideration. The Aubry operator~\shortcite{aubry2014fast} utilizes fast local Laplacian filters to conduct edge-aware multiscale manipulations.
Each operator produces a set of tone-mapped LDR as the target LDR for training. The default parameters as claimed in the papers are adopted in our experiments. Note that we train a singe model to learn all these tone mapping operators, where each individual style is incorporate in a slight-weighted style modulator.

\paragraph{Training Details.}
Our model is trained on a PC with four NVIDIA GTX 1080Ti GPU, using the Adam optimizer~\cite{kingma2014adam} with $\beta_1$=0.9 and  $\beta_2$=0.999. The batch size is set to 8. The learning rate is set to $2e{-4}$ initially and linearly decreases to $2e-6$ by 300 epochs.

\section{Results and Discussion}
\label{sec:result_discussion}

\begin{figure*}[!htbp]
	\centering
	\includegraphics[width=.95\textwidth]{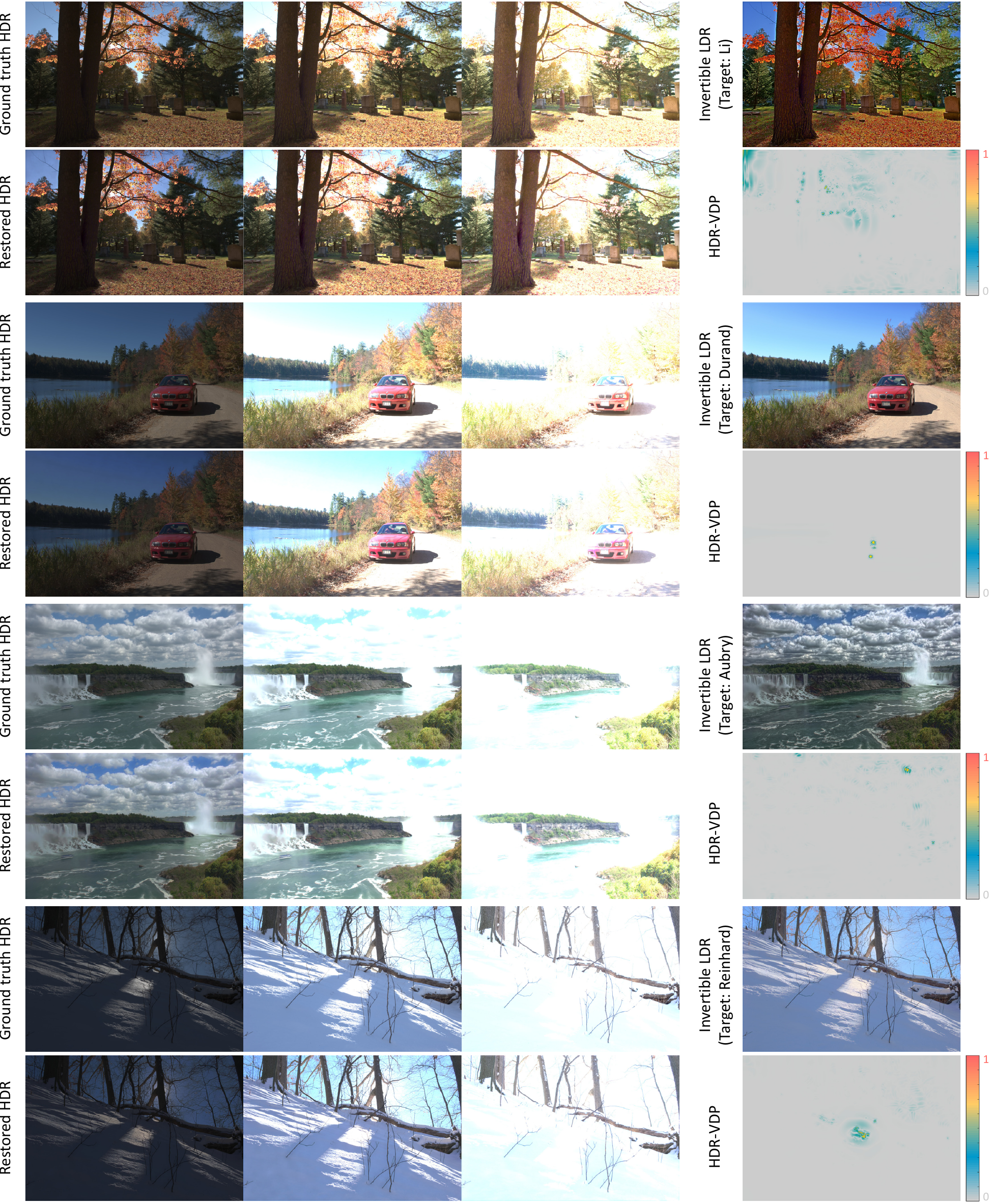}
	\vspace{-0.5em}
	\caption{Result gallery. Four target tone mapping styles are mimicked, and the corresponding HDR is restored in each case.  For each tone mapping style, the ground truth and the restored HDR images are compared in three exposure levels. The HDR-VDP maps between them are also provided on the right column. The corresponding invertible LDRs are shown on the right column (lower subfigure). The target tone mapping operator is also labeled  besides each example. }
	\label{fig:hdr_showcase}
\end{figure*}

Fig~\ref{fig:hdr_showcase} showcases our results mimicking four tone mapping styles mentioned above. The corresponding target tone mapping operator is labeled along with the sub-figure on the rightmost column.
For each case, we first generate the invertible LDR with our encoding network and quantize it to 8-bit per channel to simulate the real-world applications.  Then, we feed the quantized invertible LDR to our decoding network to restore the HDR. 
The leftmost three columns compare the original (upper row) and the restored (lower row) HDRs in three exposure levels, as we cannot visualize the HDR on paper. We can see our restored HDR can nicely restore the pixel values back to their original dynamic range. To visualize their difference, we measure the visual difference between the original and restored HDRs, by computing the HDR visual difference predictor (HDR-VDP-2.2)~\cite{narwaria2015hdr}, and visualizing them in the lower subfigure of each case, on the rightmost column. The VDP is computed in the dynamic range of the original HDR. From these VDP maps, our restored HDR are mostly accurately except in the over-exposed regions (over-exposed in the original HDR).

\begin{table}[!t]
	\centering
	\small
	\renewcommand{\tabcolsep}{5pt}
	\caption{Comparative evaluation on HDR accuracy. Four methods recovering HDR from true LDR are compared.}\vspace{-0.5em}
	\begin{tabular}{ccccc}
		\hline
		Method		                                & LDR Style     & HDR-VDP   & PU-PSNR    & PU-SSIM        \\
		\hline
		HDR Companding					   & Li         	       & 72.70		      & 53.01        & 0.9962 \\
															\hline
		\multirow{4}{*}{ExpandNet}  & Li         			& 67.53          & 41.07	 	  &  0.9642  \\ 
			    											& Reinhard       & 67.87	  	   & 44.49	        & 0.9774 \\
	    					               				    & Durand          & 70.04          & 48.71       	&  0.9882  \\ 
	    					               				    & Aubry             & 65.98         & 42.94          &  0.9594   \\
	    					               				    \hline
	    \multirow{4}{*}{HDRCNN}     & Li         		    & 67.88		       & 40.13		    & 0.9624   \\ 
	    	    											& Reinhard       & 67.32			& 38.74			& 0.9555   \\
	    	    					               			& Durand          & 72.38		    & 45.031	   & 0.9753   \\ 
	    	    					               			& Aubry            & 65.85		     & 41.28		  & 0.9533    \\
	    	    					               			\hline
	    \multirow{4}{*}{Ours}             & Li         		     & 73.43             &  52.46	   & 0.9971   \\
	    													& Reinhard       & 74.17	         &  52.93      & 0.9966  \\
	    	    	    					               	& Durand          & 75.81	         &  54.90       &  0.9980  \\ 
	    	    	    					               	& Aubry            & 72.37		       &  53.60	      & 0.9976    \\  \hline
	\end{tabular}
	\label{tab:hdr_accuracy}\vspace{-0.5em}
\end{table}

\subsection{HDR Restoration Accuracy}
\label{subsec:hdr_restoration}

%
\begin{figure*}[!t]
	\centering
	\includegraphics[width=1\linewidth]{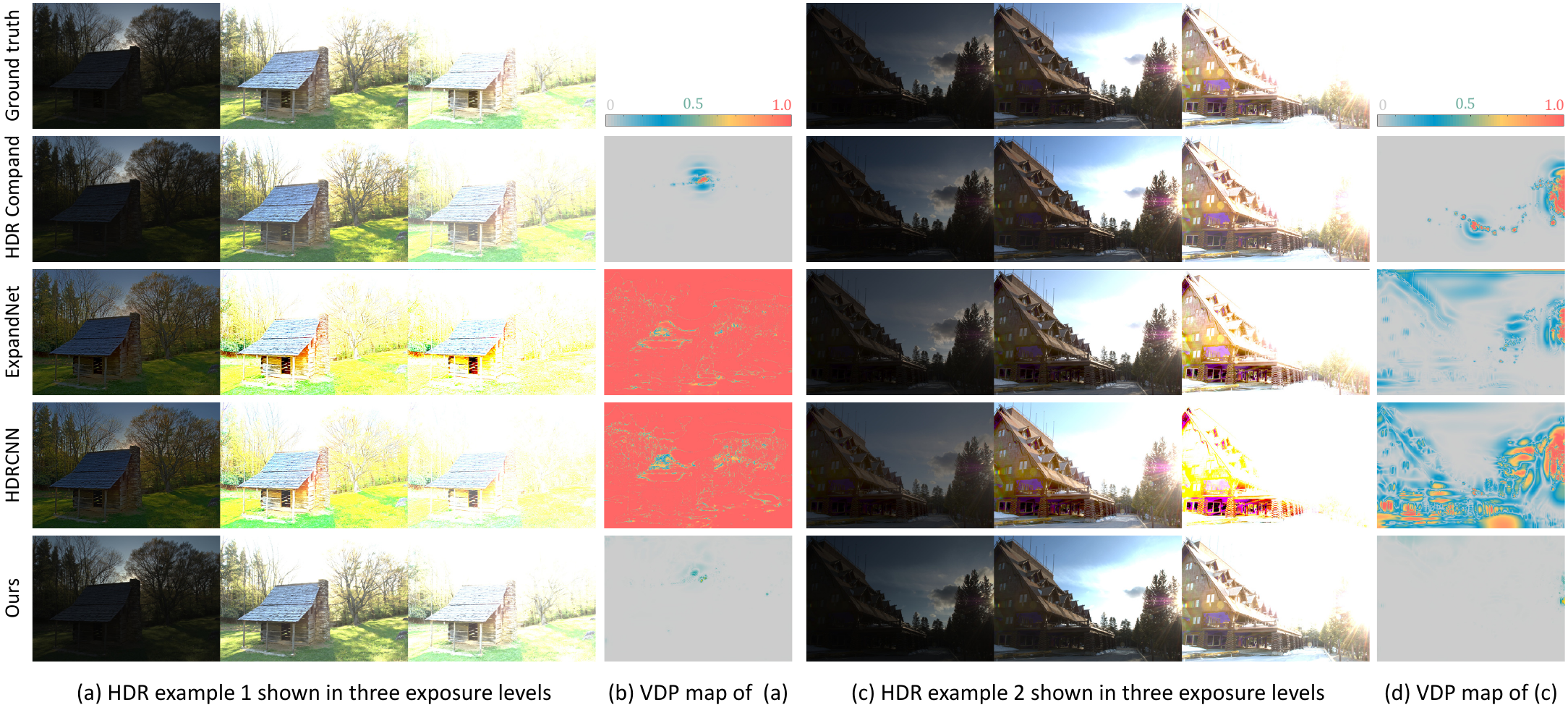}
	\caption{HDR restoration comparison among different methods. The region marked with boxes indicates detail loss and color shift. Note that the results of example 1 are based on the LDR style of Li, and the results of example 2 are based on the LDR style of Durand except for HDR Companding that can only generate fixed style (Li).}
	\label{fig:hdr_accuracy}\vspace{-0.5em}
\end{figure*}

To quantitatively evaluate the accuracy, we compute the structural similarity (SSIM), peak signal-to-noise ratio (PSNR), multi-scale structural similarity (MS-SSIM), and HDR-VDP~\cite{narwaria2015hdr} of our restored HDR images compared to the ground truth. Specifically, a perceptual uniformity (PU) encoding~\cite{aydin2008extending} is applied to adapt SSIM, PSNR, and MS-SSIM for HDR measurement. HDR-VDP provides a VDP-Q quality score for each HDR image.
We compare our restoration result to methods that recover/infer HDR from true LDR, including HDR companding~\cite{li2005compressing} and two state-of-the-art inverse tone mapping methods (HDRCNN~\cite{eilertsen2017hdr} and ExpandNet~\cite{marnerides2018expandnet}).

The quantitative results of our method and the competitors are tabulated in Table~\ref{tab:hdr_accuracy}. 
Note that HDR companding only supports its own tone mapping style. On the other hand, the learning-based inverse tone mapping methods, ExpandNet and HDRCNN, can be trained to infer HDR according to the specific tone mapping style.
Four LDR tone mapping styles, Reinhard~\shortcite{reinhard2002photographic}, Durand~\shortcite{durand2002fast}, Aubry operator~\shortcite{aubry2014fast}, and Li (HDR companding)~\shortcite{li2005compressing}, are listed separately in Table~\ref{tab:hdr_accuracy}. 
From the statistics, almost all variants of our method outperform HDR companding in the evaluated metrics. The visual results in Fig.~\ref{fig:hdr_accuracy} also confirm this. It worths noting that the variant "Ours (Li)" is trained to mimic the LDR style of HDR companding. It tells that our model can better preserve the HDR information than HDR companding even our model is mimicking the visual style of HDR companding.
Besides, our accuracy also significantly outperforms that of HDRCNN and ExpandNet, in all four tone mapping styles and in all four evaluated metrics.
Even with the state-of-the-art learning-based inverse tone mapping, there is no guarantee that the inferred HDR values confirm to the original ones. In contrast, our invertible tone mapping restores the encoded HDR information and achieve much higher fidelity.

\subsection{LDR Appearance}
\label{subsec:ldr_appearance}

%
\begin{table}
	\centering
	\renewcommand{\tabcolsep}{6pt}
	\caption{Quantitative evaluation on style simulation accuracy: tone mapping (TM) simulator~\shortcite{aubry2014fast} vs. our invertible tone mapping.}
	\begin{tabular}{ccccc}
		\hline
		\multirow{2}{*}{Tone mapping operator}  & \multicolumn{2}{c}{TM simulator}   & \multicolumn{2}{c}{Ours} \\ \cline{2-5}
		                                           & PSNR          & SSIM       & PSNR          & SSIM \\
		\hline
		Li 										  & 18.77          & 0.8275    & 27.28          & 0.9624    \\
	    Reinhard	                       & 19.84          & 0.8918    & 28.12          & 0.9701    \\
		Durand	                            & 21.38          & 0.8951    & 32.24          & 0.9826    \\
		Aubry	                              & 18.20          & 0.8448    & 26.14      	  & 0.9642    \\ 
		\hline
	\end{tabular}
	\label{table:ldr_similarity}
\end{table}

\begin{figure*}[!t]
	\centering
	\includegraphics[width=1\linewidth]{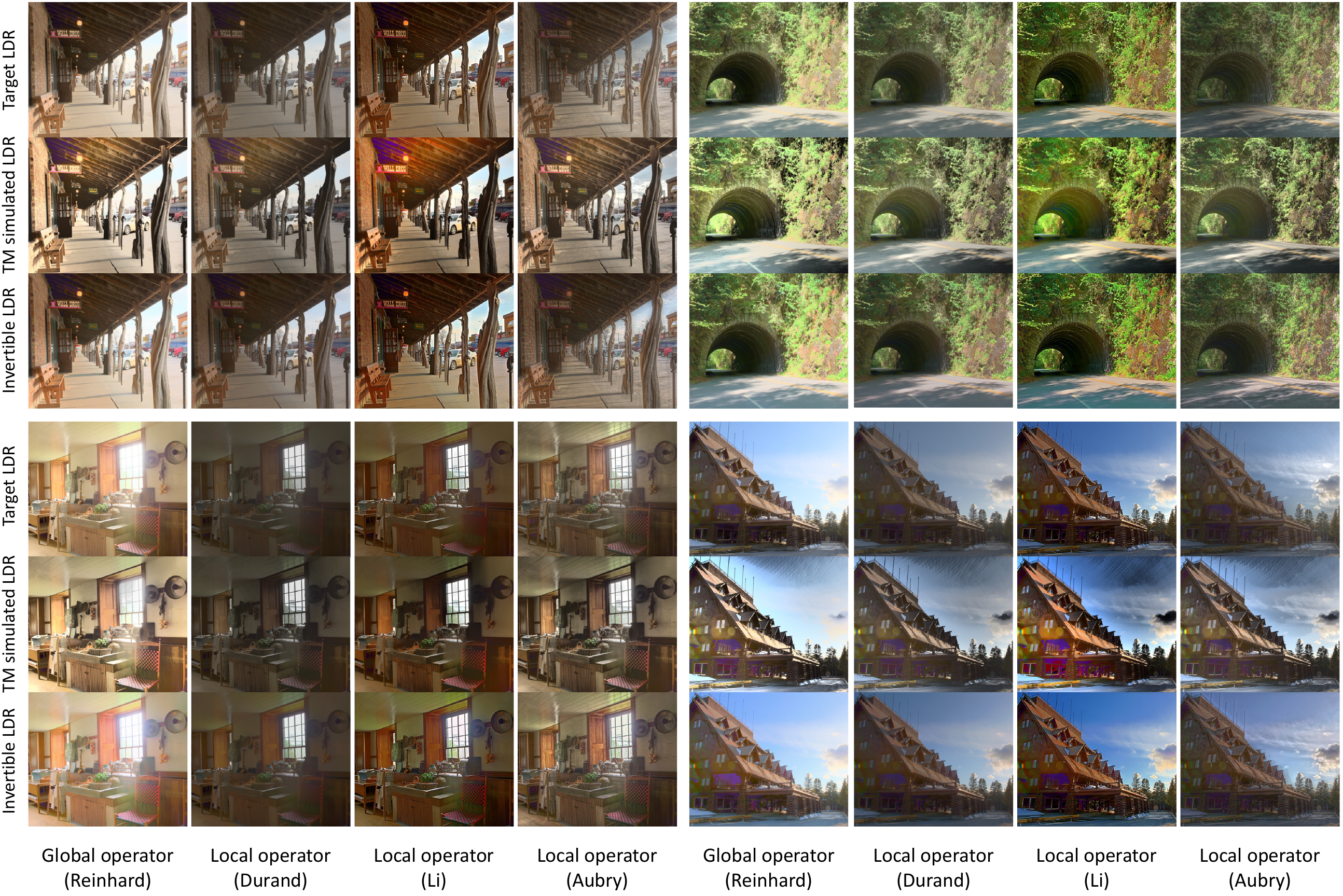}\vspace{-0.5em}
	\caption{Tone mapping with multiple styles. These results are generated by our single model with multiple style modulators.}
	\label{fig:ldr_similarity}\vspace{-0.5em}
\end{figure*}

We evaluate the LDR quality by measuring the similarity between our generated invertible LDR and the target LDR. 
For reference, the state-of-the-art generic tone mapping simulation method~\cite{aubry2014fast} is compared, which tone-map the input HDR into 8-bit LDR by simulating the style of specified tone mapping operator.
Four previously mentioned tone mapping styles are evaluated. Fig.~\ref{fig:ldr_similarity} compares the visual similarity between the simulated LDRs and the target LDRs. It can be observed that our invertible LDRs achieve high style fidelity in both global contrast distribution and local details. Nevertheless, due to the information embedding requirements, there are still certain perceptible tonal deviations from the target LDRs. Anyhow, our results still obviously outperform the tone mapping simulator, even though our invertible tone mapping additionally considers the restorability to the original HDR.
Table~\ref{table:ldr_similarity} shows the quantitative evaluation on style similarity. The high SSIM values just conform to the visual impression that our LDRs are close to the targets.
Note that the statistics in Tables~\ref{tab:hdr_accuracy} \& \ref{table:ldr_similarity} are obtained from our models trained with the default hyper parameter value $\lambda=0.5$ in Eq.~\ref{eqn:total_loss}. By decreasing  $\lambda$, we can promote the style similarity to targets at the cost of HDR restoration accuracy. So, the balance of an optimal $\lambda$ depends on the requirements of individual applications.

%

\subsection{Ablation Study}
\label{subsec:ablation_study}

We conduct ablation studies to verify the effectiveness of our proposed designs, including network structure, loss function, and training scheme. Here, to avoid defocusing, we only present the result of Durand style and make analysis based on it.

\begin{table}[!t]
	\vspace{0.1in}
	\centering
	\renewcommand{\tabcolsep}{1pt}
	\renewcommand\arraystretch{1.1}
	\caption{Quantitative evaluation on ablation studies. The LDR evaluation is on Durand operator.}
	\vspace{-0.1in}
	\begin{tabular}{cccccc}
		\hline
		\multirow{2}{*}{Model Variant}  & \multicolumn{2}{c}{Invertible LDR}   & \multicolumn{3}{c}{Restored HDR}  \\ \cline{2-6}
		                                                    & PSNR       & SSIM 	 & HDR-VDP    & PU-PSNR     &  PU-SSIM \\ \hline
		w/o global branch	                 & 30.604 	 & 0.9808   &	70.93         & 52.45           & 0.9947   \\
		\hline
		$L_2$ style loss       		             & 32.43    & 0.9770      & 74.96          & 55.06           & 0.9967   \\
		\hline
		Separate training	               	   & 30.10 	     & 0.9810    & 76.04          & 54.58          & 0.9978   \\
		Incremental training	             & 30.88       & 0.9725    & 73.34          & 51.21           & 0.9946   \\
		\hline
		Proposed method			            &  32.24       & 0.9826    & 74.18         &	54.90        &  0.9980 \\ \hline
	\end{tabular}
	\label{table:ablation_study}
\end{table}

\begin{figure}
	\centering
	\includegraphics[width=1\linewidth]{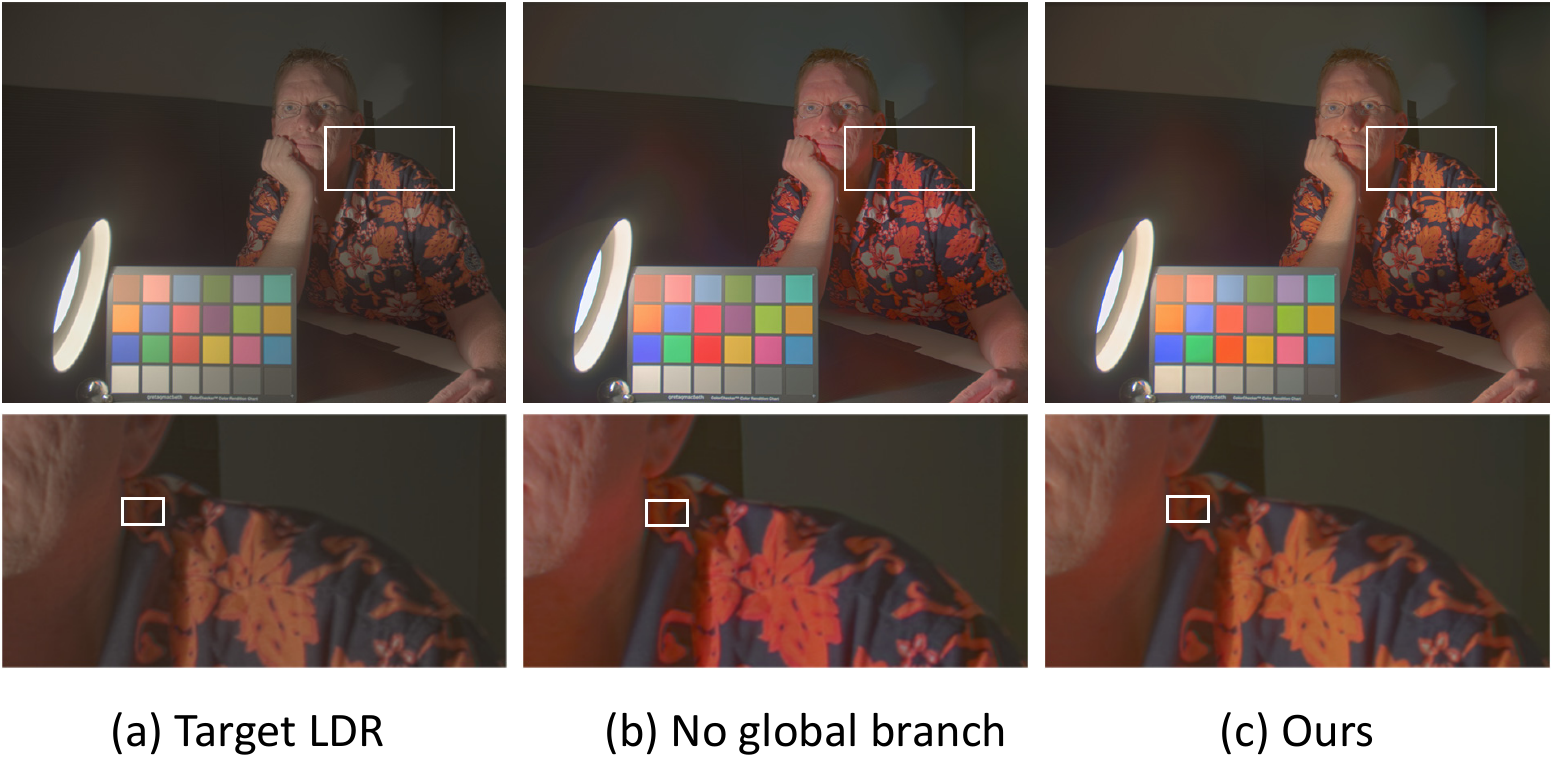}
	\caption{Visual comparison on LDRs that are respectively generated by our model variants with and without global branch. }
	\label{fig:ablation_globalbranch}
\end{figure}

\paragraph{Global Branch.}
We study the effect of the global branch design by comparing two model variants: our proposed model and our model with the global branch removed. Both variants share the same training setting to avoid other possible interferences. Fig.~\ref{fig:ablation_globalbranch}(b)\&(c) compares the generated LDRs, which tells that ablating the global branch cause noticeably color deviation with target style. The probable reason is that no global branch means a much smaller receptive field of the network, which limits the information embedding space and also impedes the global information to be used in local feature transformation. In contrast, our full model suffers less color deviation from the target LDR thanks to the global information guidance for tone mapping. In addition, the quantitative results in Table~\ref{table:ablation_study} exhibits the advantages of using global branch in promoting LDR visual quality and the HDR restoration accuracy. 

\begin{figure}
	\centering
	\includegraphics[width=1\linewidth]{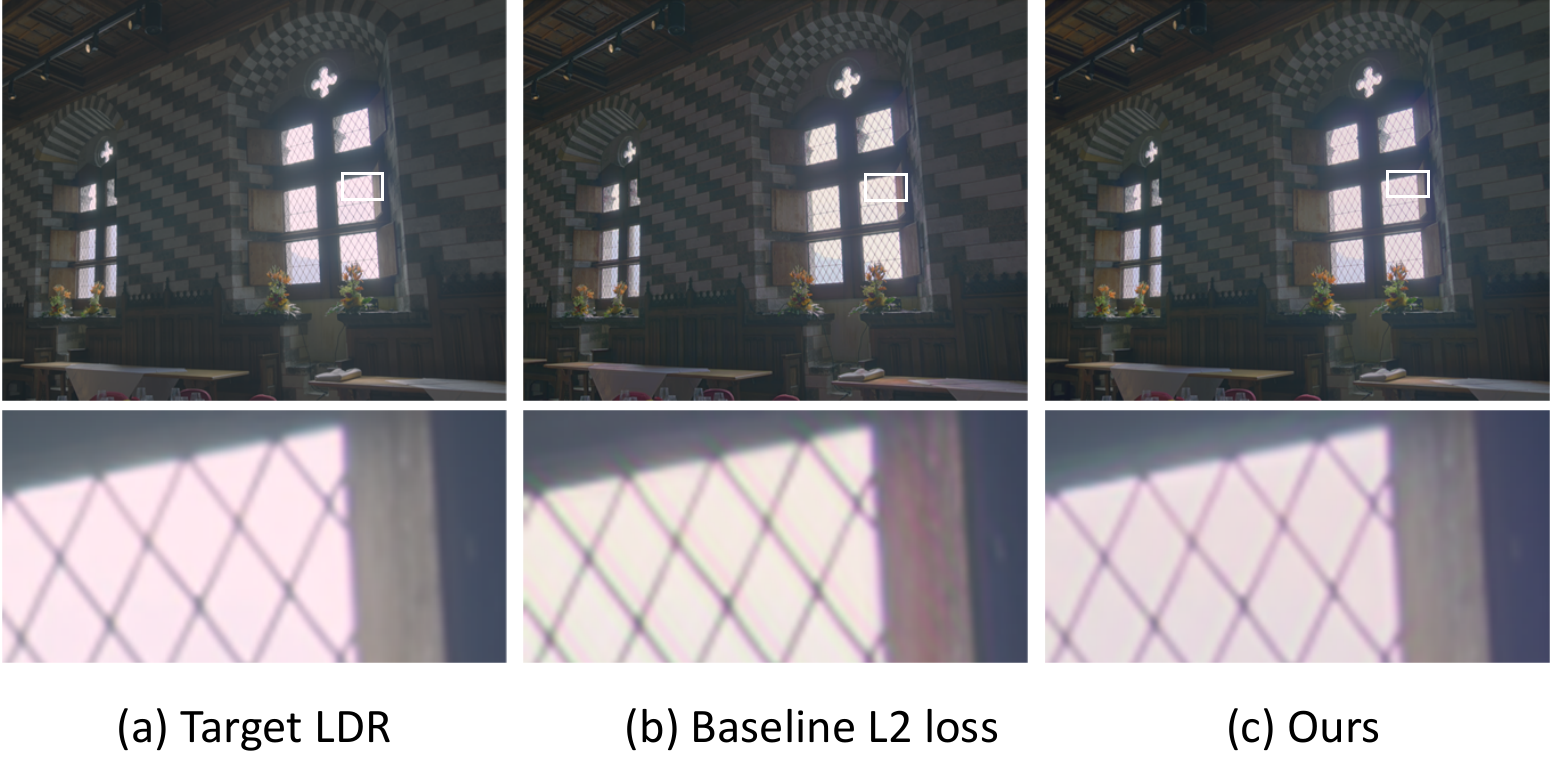}
	\caption{Visual comparison on LDRs that are generated by our model trained with $L_2$ style loss and trained with our proposed loss function. }
	\label{fig:ablation_lossfunc}
\end{figure}

\paragraph{LDR Appearance Loss.}
We study the necessity of using our style loss compared with common used $L_2$ loss only. Although the quantitative results (in Table~\ref{table:ablation_study}) measured by PSRN and SSIM shows the better performance of $L_2$ style loss, we find that the perceptual quality degrades indeed. Fig.~\ref{fig:ablation_lossfunc} evidences such an example. With only $L_2$ loss, the generated LDR suffers from color deviation and certain ghosting effects, which may be triggered by the information embedding target (see the discussion in Section~\ref{subsec:discussion}). Favorably, the additionally adopted gradient loss can suppress those double edged ghosting effect, and the perceptual loss removes abnormal color deviations.

\begin{figure}[!t]
	\centering
   	\includegraphics[width=\linewidth]{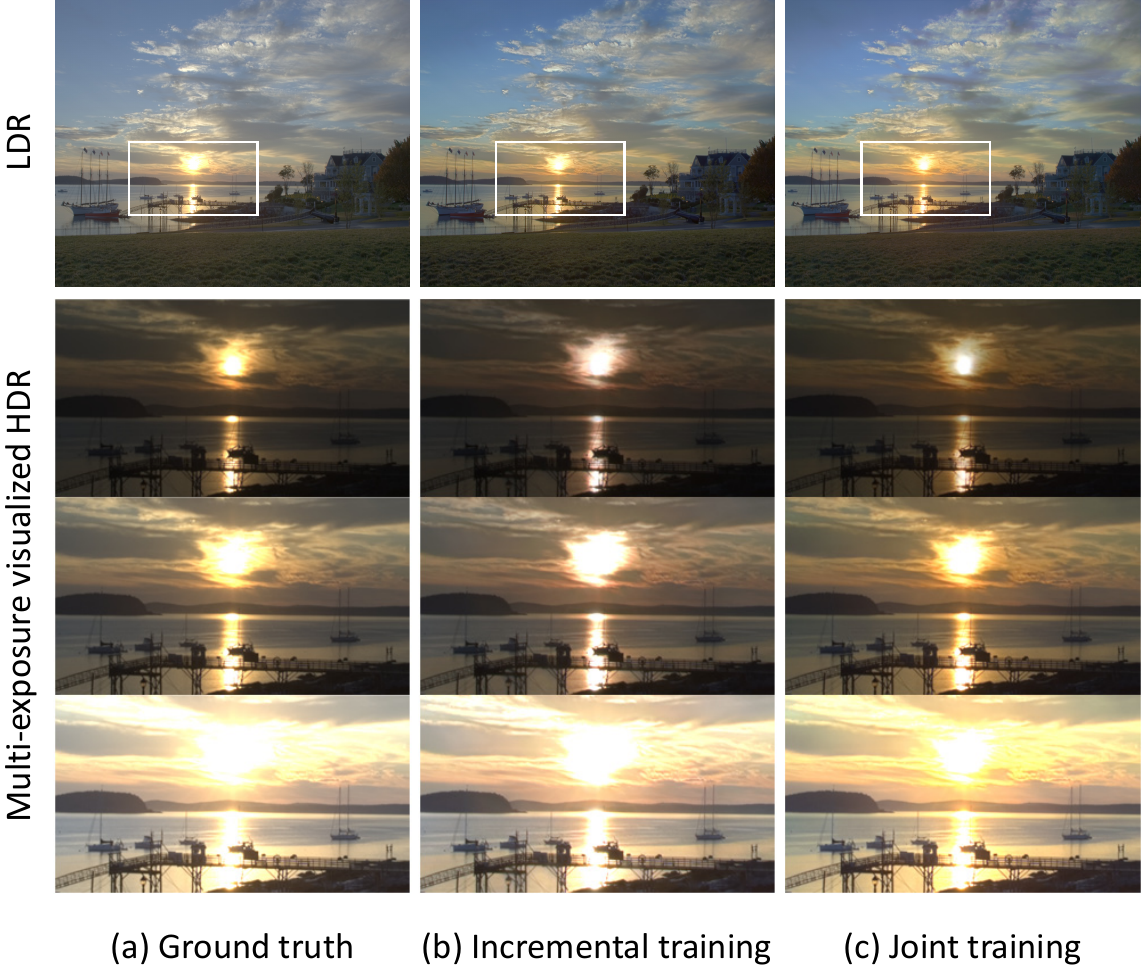}
	\vspace{-0.5em}
	\caption{Qualitative comparison on LDR similarity and HDR restoration accuracy of different training scheme. }
	\label{fig:ablation_incretrain}\vspace{-0.5em}
\end{figure}

\paragraph{Multi-style Joint Training.}
We train all the mentioned tone mapping styles jointly through a single model, which is the optimal scheme to achieve good overall performance. However, incremental training, i.e. inserting a new style modulator to the pretained model through fine-tuning, is more flexible to practical scenarios. So, we compare these two training schemes: (a) jointly train the four styles; (b) train the fourth style (i.e. Durand) along with the pretrained model that has three trained style modulators already. We evaluate the performance by measuring the accuracy of the learned fourth style.
Table~\ref{table:ablation_study} compares the quantitative results, and Fig.~\ref{fig:ablation_incretrain} offers visual inspection. 
Impressively, the incremental training only compromises the restoration accuracy slightly, which shows good practical merits.
In addition, we also present the result of training the whole model for Durand only, termed "separate training". It shows that comparing to training each style with a individual model, jointly train all the styles in a single model decreases HDR restoration accuracy while gains in LDR similarity. It is interpretable because the decoder are shared by all the styles in joint training scheme, which inevitable reduces the restoration flexibility for each specific style.

\subsection{Compression Efficiency}
\label{subsec:comp_HDR_compress}

%
 \begin{figure}[!t]
 	\centering
 	\includegraphics[width=1.0\linewidth]{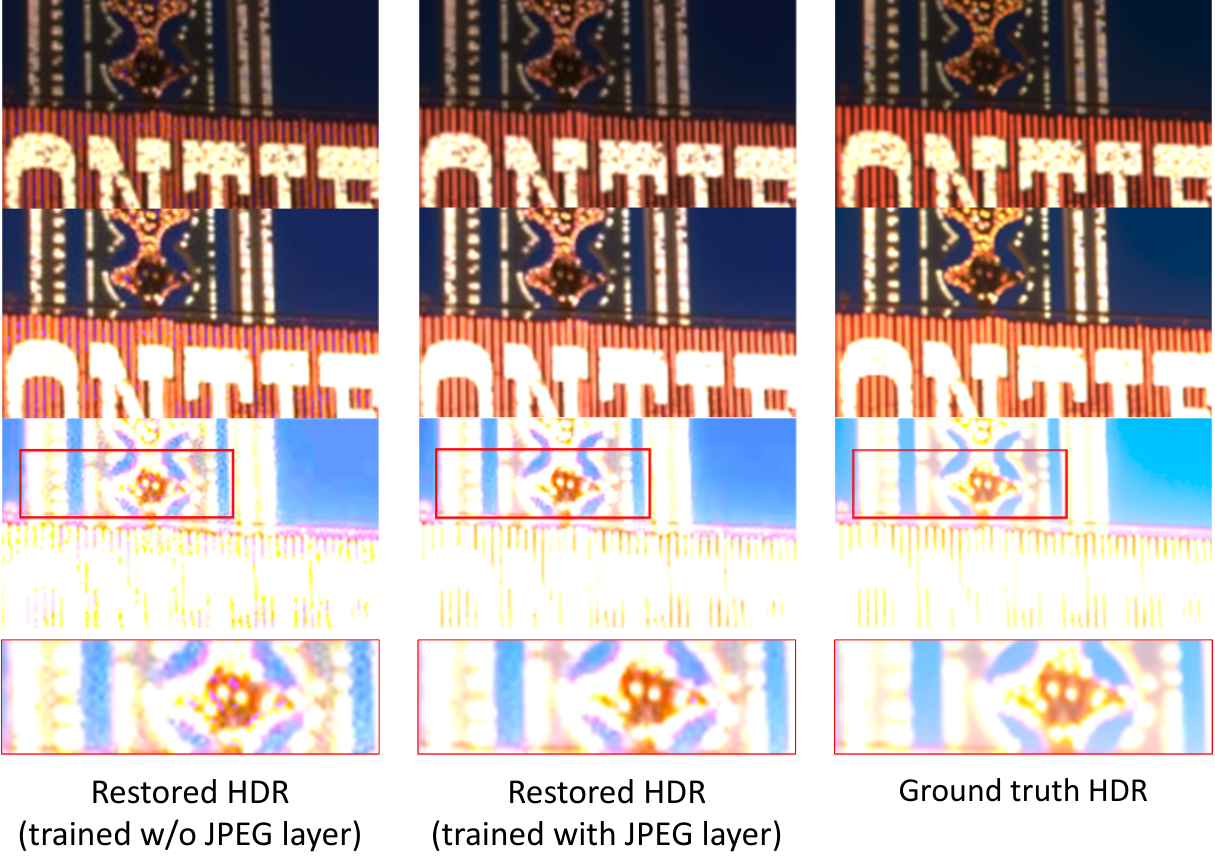}\vspace{-0.5em}
 	\caption{Ablation analysis on JPEG layer: the visual accuracy of  the restored HDRs (visualized in 3 exposure levels). The actual compression ratio for the invertible LDR of this example is around $26$.}
 	\label{fig:ablation_jpeg}\vspace{-1.0em}
 \end{figure}

\begin{figure*}[!t]
	\centering
   	\includegraphics[width=\linewidth]{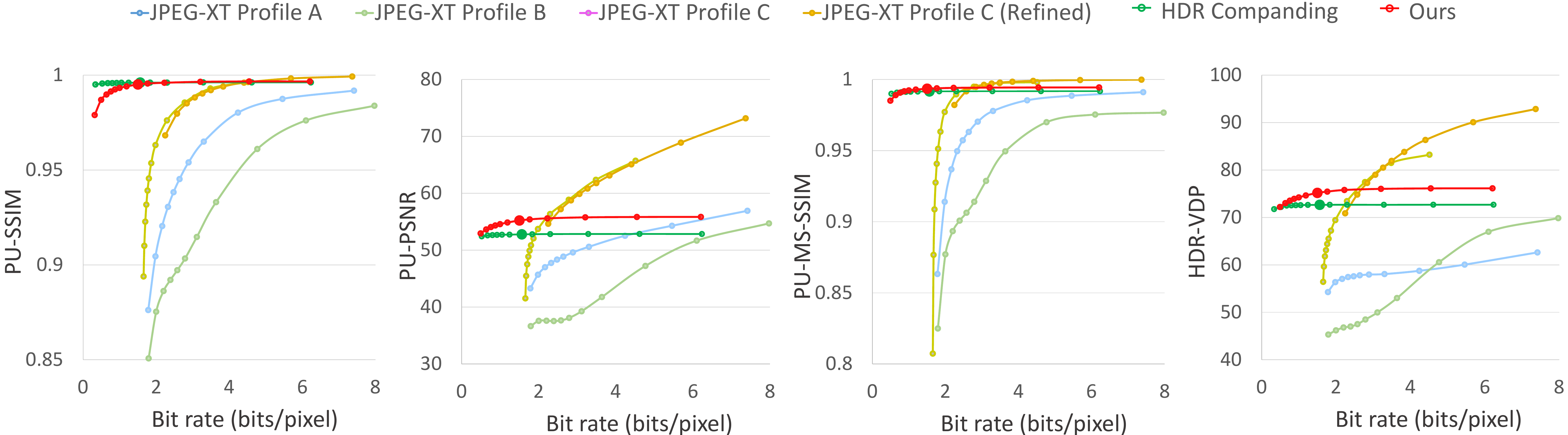}
	\vspace{-1.5em}
	\caption{Compression efficiency of different HDR compact representation methods. For JPEG-XT, the LDR layer is compressed by JPEG with quality factor $80$ and the extension layer is changed. On the curves of HDR companding and ours, the big red/green dots indicate the case of using JPEG compression quality factor $80$. }
	\label{fig:cmp_jpegxt_q80}\vspace{-1.0em}
\end{figure*}

\paragraph{JPEG Layer. }
In some sense, our invertible LDR can be use as a compact representation of HDR, which allows compatible display and distribution as a normal LDR image. To further promote its compression efficiency, we can alternatively store the invertible LDR in lossy compression format, e.g. JPEG. To this end, we only need to replace the quantization layer with a JPEG layer during training. Specifically, the JPEG layer introduce JPEG noise by performing discrete cosine transform (DCT), quantization of DCT coefficients, and inverse discrete cosine transform (IDCT). Similar to quantization layer, the non-differentiable obstacle is addressed through Straight-Through Estimator~\cite{Bengio2013}.
In Fig.~\ref{fig:ablation_jpeg}, we qualitative evaluate the restored HDR from JPEG compressed invertible LDRs that are generated by our model trained with and without the JPEG layer respectively. It shows that our model trained with JPEG layer has a good tolerance to JPEG compression noise.

 In HDR compression field, double-layer HDR compression manages to achieve backward compatibility by splitting the HDR information into an LDR image and an auxiliary data. To store the auxiliary data properly, special extension need be made to existing image format.
As the latest standard of double-layer HDR compression, JPEG-XT~\cite{artusi2016jpeg} of multiple profiles are adopted to compare with our proposed invertible LDR.
Following the practice in~\cite{artusi2016jpeg}, we plot the Rate-Distortion (R-D) curves of JPEG-XT in 2D graph by fixing the LDR layer and continuously changing the quality factor of the extension layer to obtain different bit rates. The rationale is that the accuracy of the recovered HDR is mainly determined by the auxiliary data quality in the extension layer.
Fig.~\ref{fig:cmp_jpegxt_q80} illustrates the R-D curves of JPEG-XT, HDR companding, and our method respectively. In particular, we take the target LDR (i.e. Li) as the LDR layer of JPEG-XT and compress it by JPEG with quality factor $80$. To obtain different bit rates, the LDRs of HDR companding and ours are compressed with various JPEG quality factors.
The recovered HDR is evaluated by several metrics including,  PU-SSIM, PU-PSNR, PU-MS-SSIM, and HDR-VDP. Complete comparison is available in the supplementary material.

From the results, we find that our method outperforms HDR companding in almost all cases, because our restoration accuracy is higher than HDR companding (as shown in Table~\ref{tab:hdr_accuracy}).
Besides, our invertible LDR is superior to all the profiles of JPEG-XT in low bit rates because our representation has only a singe LDR while JPEG-XT has an additional extension layer. However, at high bit rates, our results achieve comparable performance with most JPEG-XT profiles, while inferior than one to two JPEG-XT profiles. Our performance is bounded and does not improve as the bit rate increases. This is because, for any instance of our trained model, the HDR restoration accuracy is already bounded by the training parameter $\lambda$.  After certain point,  further increasing the bit rate of the JPEG compression cannot reduce the error injected by the trained model.

\subsection{Discussion}
\label{subsec:discussion}

\begin{figure}[!t]
	\centering
	\includegraphics[width=1\linewidth]{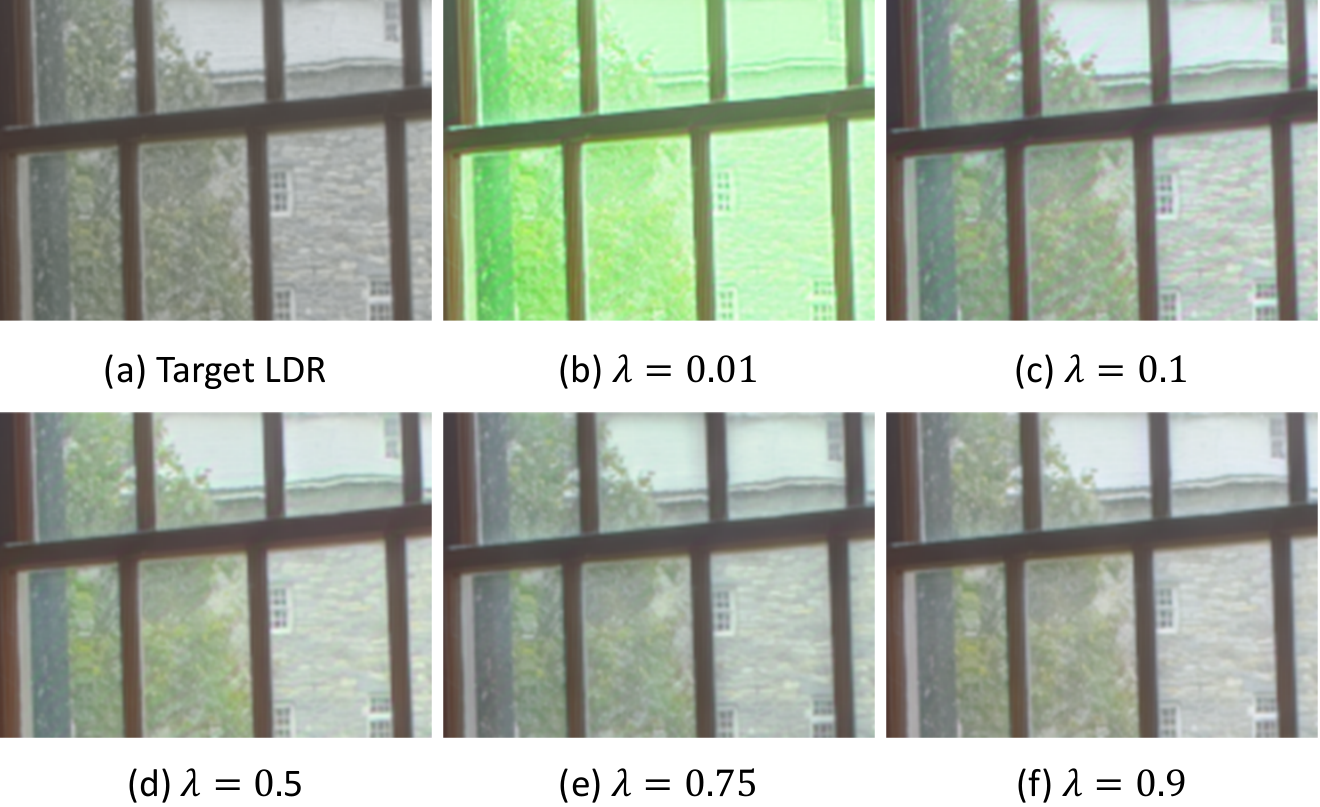}\vspace{-0.5em}
	\caption{LDR appearance change via tuning the training hype-parameter $\lambda$. Readers are recommended to inspect the color patterns on (c) and (d). }
	\label{fig:info_representation}
\end{figure}

\paragraph{Information Representation.}
The magic to our invertible LDR lies in the fact that the original HDR information is implicitly embedded among pixel values. In theory, the invertible LDR appearance shows the balance between style similarity and HDR restoration accuracy. Specifically, we can modify the balance by using a different value for hype-parameter $\lambda$ of Eq.~\ref{eqn:total_loss}. To inspect how the HDR information is represented on the invertible LDR, we trained several model variants that are associated with $\lambda=\{0.01, 0.1, 0.5, 0.75, 0.9\}$.  Their visual differences on a carefully cropped image patch are compared in Fig.~\ref{fig:info_representation}. 
When the LDR appearance constraint is loose (i.e. $\lambda=0.01$) the information embedding scheme tends to utilize both global color change and local gradient variation. However, as the appearance constraint increases, the global color change disappears while the local gradient patterns still retains for information representation (see Fig.~\ref{fig:info_representation}(b) and (c)). Anyhow, in the case of LDR appearance dominates in the training objective, the generated LDR shows very high similarity to the target LDR and visual anomaly could hardly be detected. Of course, the gain in better LDR visual quality compromises the accuracy of its HDR restoration.

\begin{figure}[!t]
	\centering
	\includegraphics[width=1\linewidth]{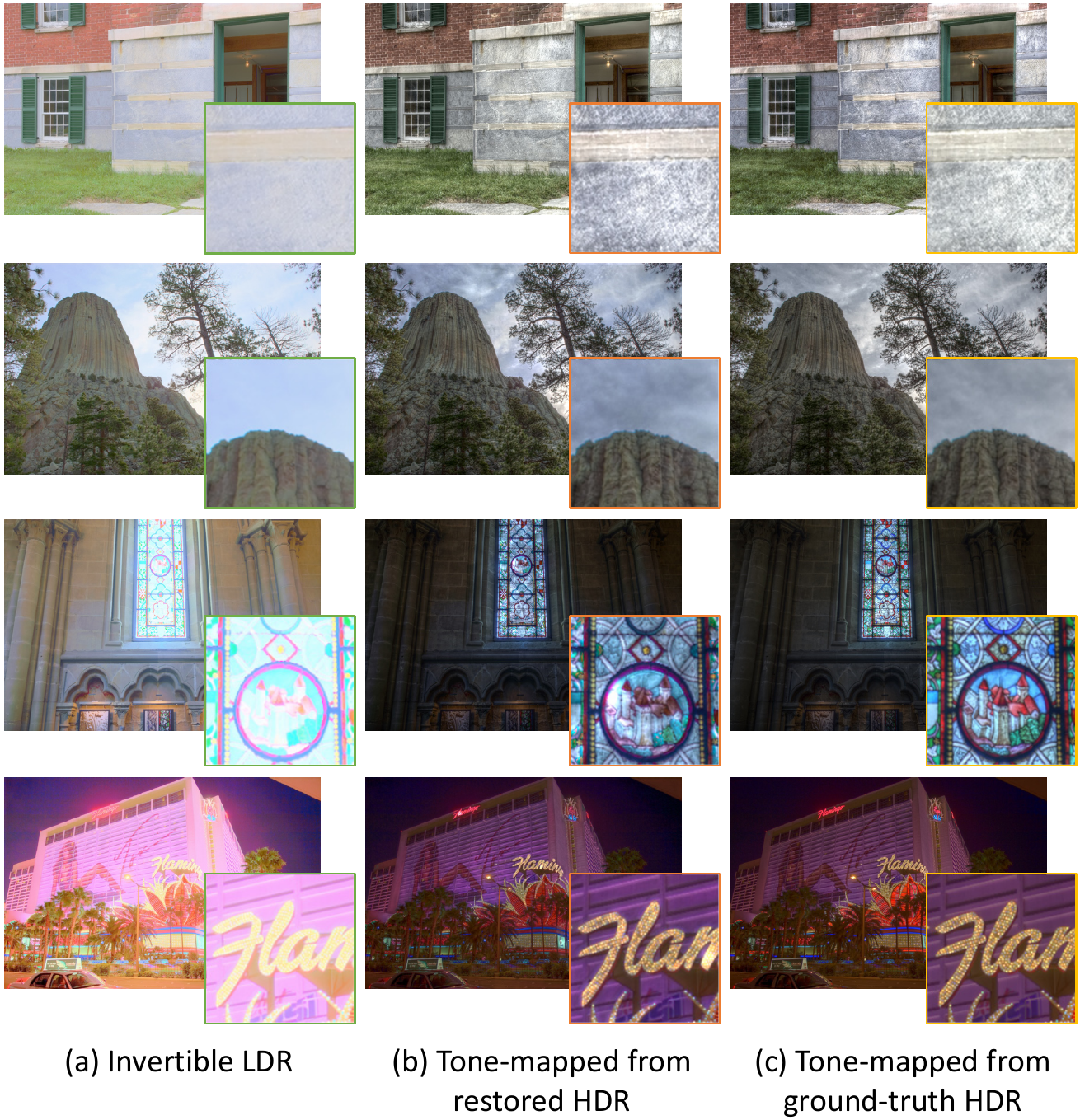}\vspace{-0.5em}
	\caption{Visual examples of re-tone-mapping from our invertible LDR. The corresponding tone-mapped LDRs from the ground-truth HDRs are provided for reference.}
	\label{fig:retone_mapp}
\end{figure}

\paragraph{Re-Tone-Mapping Applications.}
Our invertible tone mapping technique has a direct application for HDR photo release or sharing over the Internet, i.e. via Facebook, Flikir, etc. Our invertible LDR can be shared and used as normal LDR image, but it still reserves the restorability to the original HDR, which enables re-tone-mapping and editing greatly.
Fig.~\ref{fig:retone_mapp} illustrates several re-tone-mapping examples, where our invertible LDR is used to restore the original HDR and then the HDR is tone-mapped into a LDR with any of existing tone mapping operators. For reference, we also provide the LDRs tone-mapped from the ground-truth HDRs with the same parameters. We can find that the re-tone-mapped LDRs are very similar to the ones tone-mapped from the ground-truth HDRs, because our invertible LDR has a high HDR restoration accuracy (see Fig.~\ref{fig:hdr_accuracy} and Table~\ref{tab:hdr_accuracy}). Besides, some structural details invisible on the invertible LDR are accurately recovered on the re-tone-mapped images, e.g. the cloud texture and window decoration in Fig.~\ref{fig:retone_mapp}(b), which can never be achieved by existing style transfer methods.

\paragraph{Discussion on Generality.}
Our encoding network can generate invertible LDR of different styles, and they can be processed by the same decoding network for HDR restoration. Therefore, in practical application, any invertible LDR can be used to recover the HDR without the necessity to distinguish their styles. In addition, we may feed an ordinary LDR to our decoding network and expect an inferred HDR as existing inverse tone mapping methods do. It is easy to integrate this feature to our model by introducing such samples to our training process, but this naive extension is beyond the focus of this paper. In short, our proposed method has a good generality toward practical application.

\begin{figure}[!t]
	\centering
   	\includegraphics[width=\linewidth]{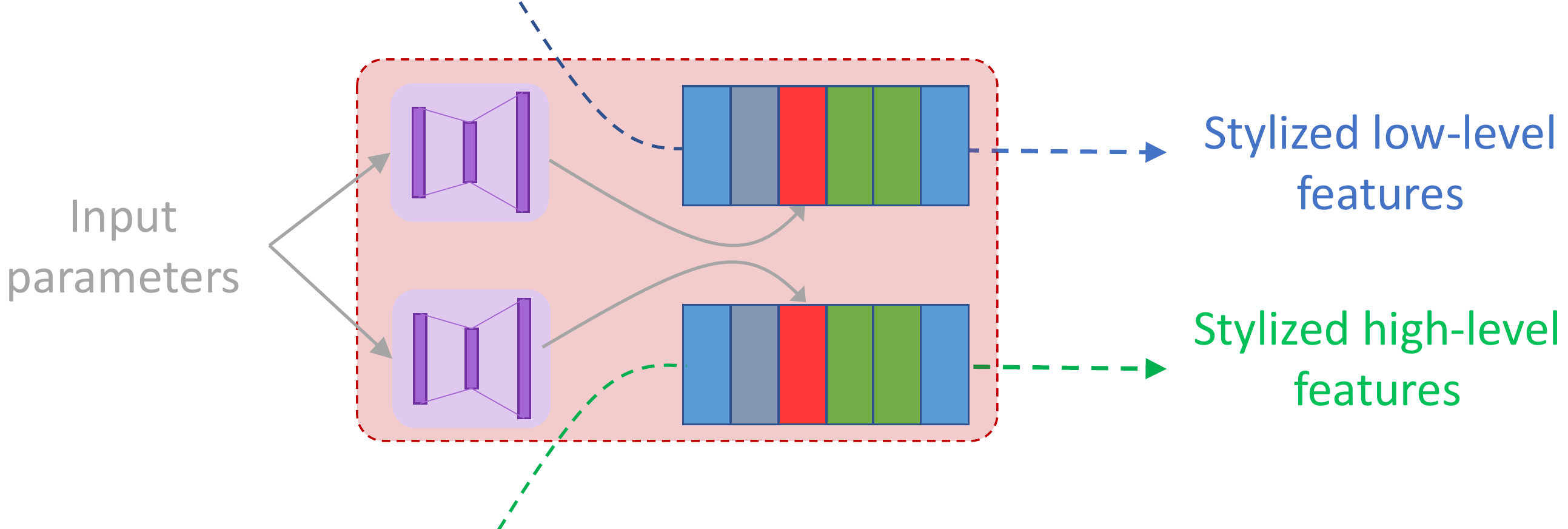}
	\vspace{-0.5em}
	\caption{Parametric style modulator. Compared basic style modulator, there are two additional fully connected networks that take as input the parameters and generate dynamic filters (labeled as red). }
	\label{fig:parametric_SM}\vspace{-0.5em}
\end{figure}

\begin{figure*}[!t]
	\centering
   	\includegraphics[width=\linewidth]{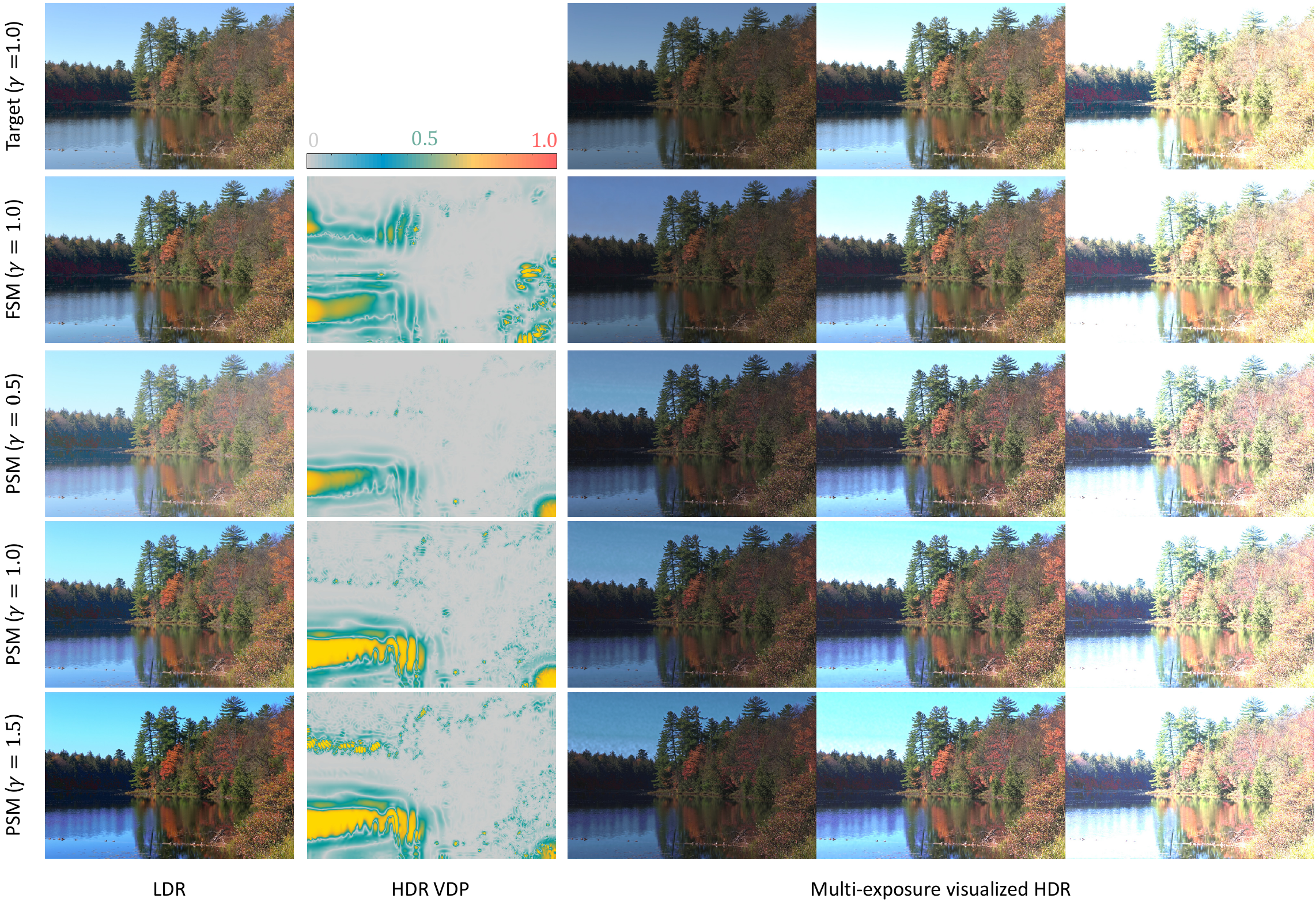}
	\vspace{-0.5em}
	\caption{Visual evaluation on parametric LDR generation. PSM denotes the model trained with parametric style modulator; FSM denotes the model trained with fixed style modulator.}
	\label{fig:evaluation_SM}\vspace{-0.5em}
\end{figure*}

\paragraph{Parametric Tone Mapping.}
Currently, the target LDR is obtained with a fixed tone mapping parameter set. However, the optimal tone mapping parameter is usually image content dependent. So, parametrized tone mapping model is more desirable in real-world applications. In fact, our proposed framework can be easily extended to support parametric style generation. Following the spirit of parametric generative model~\cite{Chen2F020}, we extend our style modulator to allow parametric style generation, as the diagram shown in Fig.~\ref{fig:parametric_SM}. To explore the effectiveness, we take Durand style as an example and utilize the parametric style modulator to model the effect of tuning gamma. The only difference to training is that the encoding network is fed with an HDR and a randomly sampled gamma parameter ($\mathcal{U}(0.5,1.5)$) for tone mapping, and the generated LDR are supervised by the parameter-dependent target LDR. Fig.~\ref{fig:evaluation_SM}  illustrates some visual comparison, which shows that the input parameter is effectively learned by the model and more importantly, decent HDR restoration accuracy is achievable by these style modulated LDRs. We leave more in-depth investigation of this direction as future work.

\subsection{Timing Statistics}
\label{subsec:timing_statistic}

We implemented our method using PyTorch~\cite{paszke2017automatic}.
All experiments were performed on a PC with an Intel Core i7-6700K @ 4.0GHz CPU and a nVidia GeForce GTX 1070 GPU.
We recorded the running times of our encoding and decoding networks for input images of different resolutions with and without GPU acceleration.
For each image resolution, we test 147 images and record the average time.
The timing statistics are shown in Table~\ref{table:timing}. The encoding and decoding speeds are symmetry due to their same network architecture.
With the GPU acceleration, our method achieves real-time speed-up to the resolution of 512$\times$512.

\begin{table}[!t]
	\centering
	\renewcommand{\tabcolsep}{3pt}
	\caption{Timing statistics (in seconds).}
	\vspace{-0.1in}
	\begin{tabular}{ccccc}
		\hline
		\multirow{2}{*}{Image Size}     & \multicolumn{2}{c}{CPU only}   & \multicolumn{2}{c}{With GPU}   \\
		\cline{2-5}  								  & Encoding   & Decoding          		& Encoding   		& Decoding	  \\ \hline
		$256 \times 256$                       & 3.889         & 3.927                     & 0.022                 & 0.020            \\ \hline
		$512 \times 512$                       & 12.917       & 13.042                   & 0.036                & 0.034           \\ \hline
		$1024 \times 1024$                   & 87.072       & 87.158                  & 0.142                 & 0.134          \\ \hline
	\end{tabular}
	\vspace{-0.1in}
	\label{table:timing}
\end{table}

\section{Conclusion}

In this paper, we proposed a novel invertible tone mapping, to convert HDR into invertible LDR that can be restored to HDR whenever necessary. To do so, we treat the tone mapping and the restoration as coupled processes, and formulate them as an encoding-and-decoding problem. Our generated invertible LDR can tolerate the noise due to JPEG compression, and its appearance can mimic the appearance of a target LDR specified by the user.
Importantly, our proposed framework supports multiple styles to be trained in a single model, where each style is represented as a light-weight style modulator. Moreover, incremental training enables us to introduce new styles to a trained model in a very efficient manner, which promotes the application potentials.
As the optimal tone mapping parameter is usually image content dependent, we  extend our model to support parametric style modulation and experimental results show the promising feasibility of this direction. Since the main focus of this paper is to demonstrate the idea of invertible tone mapping, we leave it as future work to further investigate the parametric tone mapping model.

\bibliographystyle{ACM-Reference-Format}
\bibliography{invertible_tonemap}

\end{document}